\documentclass{aa}  
\usepackage{graphicx}
\usepackage{txfonts}
\usepackage[colorlinks,breaklinks]{hyperref}
\hypersetup{linkcolor=blue,citecolor=blue,filecolor=black,urlcolor=blue}
\usepackage{pdflscape}
\usepackage{CJK}

\usepackage[T1]{fontenc}
\usepackage{fix-cm}

\usepackage{tablefootnote}

\newcommand{\FullSampleSpectra} {5028}
\newcommand{\FullSampleSNe} {3585}
\newcommand{\LCFlagSpectra} {4572}
\newcommand{\LCFlagSNe} {3280}
\newcommand{\PrepeakSpectra} {2362}
\newcommand{\PrepeakSNe} {1801}
\newcommand{\WlCoverageSpectra} {2349}
\newcommand{\WlCoverageSNe} {1793}
\newcommand{\VisibleSiSpectra} {1930}
\newcommand{\VisibleSiSNe} {1557}
\newcommand{\FailedWlCoverageInvisibleSiSpectra} {432}
\newcommand{\ResolutionCutSpectra} {660}
\newcommand{\ResolutionCutSNe} {605}
\newcommand{\SNRCutSpectra} {329}
\newcommand{\SNRCutSNe} {307}
\newcommand{\FinalSpectra} {329}
\newcommand{\FinalSNe} {307}
\newcommand{\HVFSpectra} {85}
\newcommand{\HVFSNe} {75}
\newcommand{\HVFPercentage} {26}
\newcommand{\PVFSpectra} {244}
\newcommand{\BaseBiasedSNe} {261}
\newcommand{\LowBiasSNe} {190}
\newcommand{\LowBiasSpectra} {210}
\newcommand{\LowBiasHVFSNe} {56}
\newcommand{\LowBiasHVFSpectra} {64}

\newcommand{\kms}{km~s$^{-1}$}

\newcommand{\SiII} {\ion{Si}{ii}}
\newcommand{\SiIII} {\ion{Si}{iii}}
\newcommand{\CaII} {\ion{Ca}{ii}}
\newcommand{\SII} {\ion{S}{ii}}
\newcommand{\FeII} {\ion{Fe}{ii}}
\newcommand{\CII} {\ion{C}{ii}}
\newcommand{\SiFeature} {\SiII\ $\lambda6355$}
\newcommand{\SiFeatureOther} {\SiII\ $\lambda5972$}
\newcommand{\wang} {$_\text{W}$}

\begin{document} 
\begin{CJK*}{UTF8}{gbsn}

   \title{ZTF SN~Ia DR2: High-velocity components in the \SiFeature}

    \author{L. Harvey, \inst{\ref{tcd}}
          K. Maguire, \inst{\ref{tcd}}
          U. Burgaz, \inst{\ref{tcd}}
          G. Dimitriadis, \inst{\ref{tcd}}
          J. Sollerman, \inst{\ref{okc1}}
          A. Goobar, \inst{\ref{okc2}}
          J. Johansson, \inst{\ref{okc2}}
          J. Nordin, \inst{\ref{humbolt}} \\
          M. Rigault, \inst{\ref{lyon}}
          M. Smith, \inst{\ref{lyon},\ref{lanc}}
          M. Aubert, \inst{\ref{lpc}}
          R. Cartier, \inst{\ref{udp}}
          P. Chen, \inst{\ref{weizmann}}
          M. Deckers, \inst{\ref{tcd}}
          S. Dhawan, \inst{\ref{cambridge}}
          L. Galbany, \inst{\ref{csic},\ref{ieec}}
          M. Ginolin, \inst{\ref{lyon}} \\
          W. D. Kenworthy, \inst{\ref{okc2}}
          Y.-L. Kim, \inst{\ref{lanc}}
          C. Liu (刘畅), \inst{\ref{northwestern}, \ref{ciera}}
          A.~A. Miller, \inst{\ref{northwestern}, \ref{ciera}}
          P. Rosnet, \inst{\ref{lpc}}
          R. Senzel, \inst{\ref{tcd}}
          J. H. Terwel, \inst{\ref{tcd},\ref{not-affiliation}} \\
          L. Tomasella, \inst{\ref{inaf_affliliation}}
          M. Kasliwal, \inst{\ref{caltech1}}
          R. R. Laher, \inst{\ref{ipac_affiliation}}
          J. Purdum, \inst{\ref{caltech2}}
          B. Rusholme, \inst{\ref{ipac_affiliation}}
          R. Smith, \inst{\ref{caltech2}}
          }

   \institute{School of Physics, Trinity College Dublin, College Green, Dublin 2, Ireland \label{tcd}\\
         \email{luharvey@tcd.ie}
         \and Oskar Klein Centre, Department of Astronomy, Stockholm University, SE-10691 Stockholm, Sweden \label{okc1}\
         \and Oskar Klein Centre, Department of Physics, Stockholm University, SE-10691 Stockholm, Sweden \label{okc2}\
         \and Institute of Physics, Humbolt-Universit\"at zu Berlin, Newtonstr. 15, D-12489 Berlin, Germany \label{humbolt}\
         \and Univ Lyon, Univ Claude Bernard Lyon 1, CNRS, IP2I Lyon/IN2P3, UMR 5822, F-69622, Villeurbanne, France \label{lyon}\
         \and Department of Physics, Lancaster University, Lancs LA1 4YB, UK \label{lanc}\
         \and Universit\'e Clermont Auvergne, CNRS/IN2P3, LPCA, F-63000 Clermont-Ferrand, France \label{lpc}\
         \and Instituto de Estudios Astrof\'isicos, Facultad de Ingenier\'ia y Ciencias, Universidad Diego Portales, Av. Ej\'ercito Libertador 441, Santiago, Chile \label{udp}\
         \and Department of Particle Physics and Astrophysics, Weizmann Institute of Science, Rehovot 7610001, Israel \label{weizmann}\
         \and Institute of Astronomy and Kavli Institute for Cosmology, University of Cambridge, Madingley Road, Cambridge CB3 0HA, UK \label{cambridge}\
         \and Institute of Space Sciences (ICE, CSIC), Campus UAB, Carrer de Can Magrans, s/n, E-08193, Barcelona, Spain \label{csic}\
         \and Institut d'Estudis Espacials de Catalunya (IEEC), E-08034 Barcelona, Spain \label{ieec}\
         \and Department of Physics and Astronomy, Northwestern University, 2145 Sheridan Road, Evanston, IL 60208, USA \label{northwestern}\
         \and Center for Interdisciplinary Exploration and Research in Astrophysics (CIERA), 1800 Sherman Ave., Evanston, IL 60201, USA \label{ciera}\
         \and Nordic Optical Telescope, Rambla Jos\'e Ana Fern\'andez P\'erez 7, ES-38711 Bre\~na Baja, Spain \label{not-affiliation}\
         \and INAF - Osservatorio Astronomico di Padova, Vicolo dell'Osservatorio 5, I-35122 Padova, Italy \label{inaf_affliliation}\
         \and Division of Physics, Mathematics, and Astronomy, California Institute of Technology, Pasadena, CA 91125, USA \label{caltech1}\
         \and IPAC, California Institute of Technology, 1200 E. California Blvd, Pasadena, CA 91125, USA \label{ipac_affiliation}\
         \and Caltech Optical Observatories, California Institute of Technology, Pasadena, CA 91125, USA \label{caltech2}\
         }

    \titlerunning{ZTF SN~Ia DR2: - \SiFeature\ HVFs}
    \authorrunning{L. Harvey, et al.}

   \date{Received XXX; accepted YYY}

  \abstract
    {The Zwicky Transient Facility SN Ia Data Release 2 provides a perfect opportunity to perform a thorough search for, and subsequent analysis of, high-velocity components in the \SiFeature\ feature in the pre-peak regime. The source of such features remains unclear, with potential origins in circumstellar material or density/abundance enhancements intrinsic to the SN ejecta. Therefore, they may provide clues to the elusive progenitor and explosion scenarios of SNe Ia. We employ a Markov-Chain Monte Carlo fitting method followed by Bayesian Information Criterion testing to classify single and double \SiFeature\ components in the DR2. The detection efficiency of our classification method is investigated through the fitting of simulated features, allowing us to place cuts upon spectral quality required for reliable classification. These simulations were also used to perform an analysis of the recovered parameter uncertainties and potential biases in the measurements. Within the \FinalSpectra\ spectra sample that we investigate, we identify \HVFSpectra\ spectra exhibiting \SiFeature\ HVFs. We find that HVFs decrease in strength with phase relative to their photospheric counterparts -- however, this decrease can occur at different phases for different objects. HVFs with larger velocity separations from the photosphere are seen to fade earlier leaving only the double components with smaller separations as we move towards maximum light. Our findings suggest that around three quarters of SN Ia spectra before $-11$~d show high-velocity components in the \SiFeature\, with this dropping to around one third in the six days before maximum light. We observe no difference between the populations of SNe Ia that do and do not form \SiFeature\ HVFs in terms of SALT2 light-curve parameter $x_1$, peak magnitude, decline rate, host mass, or host colour, supporting the idea that these features are ubiquitous across the SN Ia population.}


   \maketitle
%

\section{Introduction}

\label{intro}
Believed to be the thermonuclear explosion of a white dwarf due to interactions with a binary companion, Type Ia supernovae (SNe Ia) are a well-studied class of transients. With the normal SNe Ia following a strict relationship between their absolute magnitude and their light curve shape (\citealt{phillips_relation_pskovskii}; \citealt{phillips_relation_phillips}), their application as standardised candles was key to the discovery of the accelerating expansion of the Universe and in turn, dark energy (\citealt{dark_energy_riess}; \citealt{dark_energy_perlmutter}).

The observational properties and evolution of SNe Ia have been well investigated in the literature, leading to further subdivision of the class into an ever growing number of subclasses (see \citealt{taubenberger_review} for a review). Many diagnostics have been developed to explore this diversity, grouping together clusters of similar SNe Ia in various parameter spaces; many of which revolve around the \SiFeature\ absorption feature. The Branch classification scheme \citep{branch2006, branch2009} divides the SN Ia population into four classes (core-normal, shallow-silicon, broad-line, and cool) based on the pseudo-equivalent widths of the \SiFeature\ and $\lambda$5972 lines at maximum light, with the shallow-silicon and cool groups broadly aligning with the 91T- and 91bg-like subtypes respectively. \cite{wang_classification} also drew their classifications from maximum-light spectra, defining all objects with \SiFeature\ velocity below 12000~\kms\ as normal-velocity, with those above this threshold as high-velocity. Analysing not just the \SiFeature\ velocity from a single epoch, but the rate at which this velocity drops, \cite{Benetti_types} divide the population into low velocity-gradient (LVG) and high velocity-gradient (HVG), with the post peak velocity decline as less than or greater than 70~\kms\ day$^{-1}$, respectively.

While the velocities and widths of the \SiFeature\ line have been well characterised, little focus has been given to the potential presence of high-velocity features (HVFs). Seen predominantly in the \CaII\ NIR and \CaII\ H\&K absorption lines in spectra taken up to maximum light \citep{hatano1999, kasen2003, wang2003, gerardy2004, thomas2004, mazzali2005, mazzali_99ee, ptf, marion2013}, these features appear as secondary absorption components several thousands of \kms\ to the blue of the photospheric-velocity component (PV).  While less common than in the calcium, a number of SNe Ia have been seen to possess these HV components in the \SiFeature\ \citep{quimby2006, childress2013, marion2013}, with HVFs also having been reported in \SiIII, \SII, and \FeII\ lines \citep{hatano1999, marion2013}. The source of these HVFs remains uncertain, with potential origins in density or abundance enhancements in the ejecta or their formation due to circumstellar material  \citep[e.g.][]{mazzali2005,Tanaka_3d_hvf}.

Few samples of HVF spectra have been constructed and studied in the past, with any discussion of HVFs being typically conducted on an object-by-object basis. \cite{childressHVF} studied a sample of 58 low-redshift (z~$\leq$~0.03) SNe Ia with maximum-light spectra to investigate the relation between the \SiFeature\ and \CaII\ NIR features. Their analysis of the strengths of the HV components in the \CaII\ NIR feature indicated that HVF strength decreases with increasing light-curve decline rate, being absent altogether in the rapidly-evolving targets. \CaII\ NIR HVF strength was also shown to decrease with increasing silicon velocity at maximum light, with clearly `high-velocity' SNe Ia \citep[v$_{\text{Si}}$~$\geq$~12000~\kms\ at peak;][]{wang_classification} showing no HVFs. \cite{ptf} investigated a sample of 264 SNe Ia from the Palomar Transient Factory (PTF), including spectra obtained more than two weeks before maximum light. They found that HVFs in \CaII\ NIR line appear to be ubiquitous at early times, with $\sim$95 per cent of SNe Ia with a spectrum before $-$5 days from peak displaying a HV component.

\cite{silvermanHVF} conducted a search for HVFs in 445 spectra from 210 objects in the \SiFeature, \CaII\ NIR, and H\&K features. Their results agreed with \cite{childressHVF} in finding underluminous objects to lack \CaII\ NIR HVFs unlike the rest of the subclass. The less common \SiFeature\ HVFs were shown to only appear at earlier phases and are more commonly found accompanied by higher photospheric velocities. \cite{silvermanHVF} also found stronger HV components in the \SiFeature\ in objects lacking early \CII\ absorption with redder colours around peak.

This work aims to identify spectra in the Zwicky Transient Facility \citep[ZTF;][]{Bellm2019_ZTF, Graham2019_ZTF, Masci2019_ZTF, Dekany2020_ZTF} Cosmology Data Release 2 (`ZTF Cosmo DR2' or simply `DR2'; Rigault et al. in prep.) possessing HV components in the \SiFeature\ feature and then analyse the resulting distributions of phase, component velocity separation ($\Delta v$), and corresponding light curve properties, as well as investigate the correlations drawn from previous samples. In Sect.~\ref{sec:observations} we introduce the dataset, in Sect.~\ref{sec:method} we present the fitting algorithm for the spectra and our simulations, the results of the fitting to the real data and presented in Sect.~\ref{sec:results} and subsequently discussed in the context of the literature in Sect.~\ref{sec:discussion}.

\section{Observations and sample definition}

\label{sec:observations}

The ZTF Cosmo DR2 comprises the ZTF data for SNe Ia in the first three years of operations (2018 -- 2020), which is the largest SN Ia sample to date from a single untargeted survey. This dataset consists of forced photometry light curves and spectroscopy for each SN Ia. A dataset overview containing statistics and technical details concerning the photometric and spectroscopic observations, as well as details on sub-classifications and host associations, can be found in \cite{mickael_dr2}, along with a list of accompanying analysis papers written by the ZTF Ia working group.

\subsection{Data acquisition and reduction}

\label{sec:data}
The SN Ia spectra in our sample come from a variety of sources. In this section, we describe the instrument and telescope, as well as data reduction technique for each telescope and instrument setup. Our measurements are sensitive to the wavelength and relative flux calibration across the \SiFeature\ feature, but we do not make use of the absolute values so absolute flux calibration to photometric measurements is not performed. These spectra are publicly released as part of the ZTF DR2 data release, with details on how to access the spectra and metadata provided in \cite{mickael_dr2}. We perform a number of cuts on our sample to select pre-maxiumum light spectra that reliable phase estimates from their light curves and have sufficient signal to noise in the region of the \SiFeature\ feature (see Section \ref{sec:sample_def}). 

Spectra from the European Southern Observatory's (ESO) New Technology Telescope (NTT) were obtained with ESO Faint Object Spectrograph and Camera version 2  \citep[EFOSC2;][]{EFOSC} as part of the ePESSTO/ePESSTO+ collaboration \citep{smartt_pessto} at the La Silla Observatory. These spectra were reduced using a custom built pipeline described in \cite{smartt_pessto} to provide wavelength- and flux-calibrated spectra. A number of spectra come from the 2 m Liverpool Telescope  \citep[LT;][]{Steele_LT2004} using the Spectrograph for the Rapid Acquisition of Transients \citep[SPRAT;][]{SPRAT} at the Observatorio del Roque de los Muchachos. The spectra were reduced with the pipeline of \cite{Barnsley_FRODOSpec2012}, adapted for SPRAT, along with a custom Python pipeline \citep[][]{Prentice_2018cow}. 

Spectra were obtained with the SuperNova Integral Field Spectrograph \citep{SNIFS} on the University of Hawai'i 88-inch Telescope (UH88) at the Mauna Kea Observatories. The spectra were reduced using the pipeline of the Spectroscopic Classification of Astronomical Transients (SCAT) Survey \citep{Tucker_SNIFS}. This pipeline is insufficient for studies involving absolute spectrophotometric calibration. However, our measurements involve velocities and relative line fluxes of the \SiFeature\ feature and so this is not an issue. We are also not concerned with the region affected by the dichroic crossover ($\sim$5000 - 5200 \AA) that require special flat field images that are not applied by this pipeline. Spectra were also obtained at the Las Cumbres Observatory's FLOYDS Spectrograph on the Faulkes Telescope North (FTN) and on the Faulkes Telescope South (FTS) at Haleakala and Siding Spring, respectively \citep{FTN_FTS}. These spectra were reduced and calibrated using a custom pipeline\footnote{\textsc{floyds$\_$pipeline},{https://github.com/svalenti/floyds$\_$pipeline}} described in \cite{Valenti2014_Faulkes}. 

The KAST Spectrograph \citep{KAST} on the Shane 3 m Telescope at the Lick Observatory was used to obtain spectra that were reduced and calibrated using a custom Python pipeline, as detailed in \cite{Dimitriadis_2020esm}. Spectra were obtained with the Low Resolution Imaging Spectrometer \citep[LRIS;][]{LRIS1, LRIS2, LRIS3} on the Keck I telescope and reduced and calibrated using \textsc{lpipe} \citep{Perley_LRIS}. The spectrum obtained with DEIMOS on the Keck II telescope was reduced using the \textsc{pypeit} software package \citep{Prochaska_pypeIt}\footnote{https://github.com/pypeit/PypeIt}. The Double Beam Spectrograph (DBSP) on the 200-inch Telescope (P200) at Palomar Observatory was used to obtain a number of spectra, these were reduced used a PyRAF-based pipeline, `pyraf-dbsp' \citep{Bellm_DBSPreductions}.

Spectra were obtained with the Asiago Faint Objects Spectrograph and Camera (AFOSC) on the 1.82 m Copernico Telescope at the INAF Osservatorio Astronomico di Padova. They were reduced using standard tasks in \textsc{iraf}, including wavelength calibration with arc lamp spectra and flux calibration using spectrophotometric standard stars. The spectra obtained at Alhambra Faint Object Spectrograph and Camera (ALFOSC) on the Nordic Optical Telescope (NOT) at the Observatorio del Roque de los Muchachos was reduced using a custom \textsc{pypeit}  \citep{Prochaska_pypeIt} environment\footnote{https://gitlab.com/steveschulze/pypeit$\_$alfosc$\_$env}. Spectra were obtained using the Dual Imaging Spectrograph (DIS) on the Astrophysical Research Consortium 3.5m Telescope (ARC) at the Apache Point Observatory (APO) and reduced using standard routines in \textsc{iraf}, as in e.g.~\cite{Sharma_Iacsm}. Spectra obtained at the Goodman Spectrograph \citep{Goodman} on the Southern Astrophysical Research Telescope (SOAR) at Cerro Tololo Inter-American Observatory were reduced and calibrated using a custom Python pipeline, as detailed in e.g.~\cite{Dimitriadis_2020esm}. Spectra were obtained at the Very Large Telescope with the Focal Reducer and Low Dispersion Spectrograph 2 (FORS2) on UT1 of the Very Large Telescope (VLT) at the ESO Paranal Observatory. These spectra were reduced using a custom Python pipeline\footnote{\textsc{forsify}, https://github.com/afloers/forsify} based on \textsc{pypeit} \citep{Prochaska_pypeIt}. 

Spectra were observed with the Inamori-Magellan Areal Camera and Spectrograph \citep[IMACS][]{dressler11} mounted on the Magellan-Baade telescope at Las Campanas observatory. The data reduction was performed in \textsc{iraf}\footnote{IRAF was distributed by the National Optical Astronomy Observatory, which is operated by the Association of Universities for Research in Astronomy, Inc., under cooperative agreement with the National Science Foundation.} following standard reduction procedures. The Low Dispersion Survey Spectrograph 3 \citep[LDSS-3;][]{LDSS3_Magellan} on the Magellan-Clay Telescope at Las Campanas Observatory was used to obtain one spectrum, which was reduced using standard \textsc{iraf} routines as described in \cite{Hamuy_CSP} using the same procedure as described for IMACS above.

Spectra were obtained with the Optical System for Imaging and low-Intermediate Resolution Integrated Spectroscopy \citep[OSIRIS;][]{OSIRIS} on the Gran Telescopio Canarias (GTC) at the Observatorio del Roque de los Muchachos. These data were reduced and calibrated following the method in Piscarrera (in prep.) using custom routines based on \textsc{pypeit} \citep{Prochaska_pypeIt}.  Some spectra were obtained with the Device Optimized for the Low Resolution (DOLORES) on the Telescopio Nazionale Galileo (TNG) at the Observatorio del Roque de los Muchachos and reduced using \textsc{pypeit} \citep{Prochaska_pypeIt}, following \cite{Das2023_carichiib}.

Spectra were obtained with Gemini Multi-Object Spectrographs (GMOS) \citep{GMOS1, GMOS2} on the Gemini North Telescope at the Mauna Kea Observatories. The spectra were reduced using standard \textsc{iraf}/\textsc{pyraf} and Python routines, following \cite{Dimitriadis_2020esm}. The Intermediate-Dispersion Spectrograph and Imaging System (ISIS) and the Auxiliary-port Camera \citep[ACAM;][]{Benn_ACAM} on the William Herschel Telescope (WHT) at the Observatorio del Roque de los Muchachos were used to obtain spectra. These were reduced and calibrated using standard \textsc{iraf} routines.  The Robert Stobie Spectrograph (RSS) \citep{RSS} on the South African Large Telescope (SALT) \citep{SALT} at the South African Astronomical Observatory (SAAO) was used to obtain spectra. The spectra were reduced using the custom pipeline, PySALT \citep{PYSALT} to produce wavelength- and flux-calibrated spectra. Spectra were obtained with the DeVeny spectrograph on the 4.3 m Discovery Channel Telescope (DCT), which was reduced using standard \textsc{iraf} routines, including wavelength and flux calibration \citep{2017ApJ...842...29H}. The Spectral Energy Distribution Machine \citep[SEDm;][]{SEDM, SEDM_Young-Lo, SEDM_Jeremy} on the P60 \citep{Cenko2002_P60} was used to obtain spectra, which were subsequently reduced using \textsc{pysedm} \citep{SEDM_Mickael}.

\color{black}
We do not require absolute flux calibration for our measurements but our measurements are impacted by the relative flux and the wavelength calibration.  Not all the spectral reduction techniques used for our sample provide meaningful flux uncertainties. Therefore, as described in Section \ref{sec:method_spectral_line_fitting}, we estimate the uncertainty on the flux using the standard deviation of the values in certain continuum regions near to the feature of interest. For SN Ia spectra, where flux uncertainties were available, we cross-checked against the uncertainty estimates from the standard deviation near the feature of interest and found them to be consistent.  In Section \ref{sec:sample_def}, we describe how we performed a cut on the sample based on the signal to noise in the \SiFeature\ region. Each remaining spectrum after this cut was inspected manually while choosing the continuum regions and if any cosmic rays or host galaxy lines were identified they were masked out prior to fitting the \SiFeature\ feature. We include an uncertainty in our fitting of 200~\kms\ to account for additional velocity offsets associated with motions of the SNe Ia in their host galaxies.

\subsection{Sample definition}
\label{sec:sample_def}
The starting point for defining our sample is the subset of the 3628 confirmed SNe Ia in the DR2 with spectra from the aforementioned facilities for which sufficient data information and reliable reductions could be performed, amounting to \FullSampleSNe\ targets with \FullSampleSpectra\ spectra. 43 SNe Ia that were included in the full DR2 data release are excluded from our sample due to a lack of sufficient information on the observations and data reduction techniques. As our analysis will be restricted to spectra in the pre-peak regime, we require sufficient photometry to produce a reliable estimate of spectral phase. We used the suggested cuts of \cite{mickael_dr2} on the SALT2 light-curve fitter \citep{guy2007} outputs, of fit probability greater than 10$^{-5}$ and uncertainties smaller than the quoted values for the light curve width parameter, $x_1$ ($\delta x_1<1$), color parameter, $c$ ($\delta c<0.1$), and the time of maximum light, $t_0$ ($\delta t_0<1$ d). We do not impose any constraints on the measured values of $x_1$ or $c$.

\textcolor{black}{To maximise our spectral sample, we investigated all SNe Ia that failed to meet these criteria in order to avoid cutting otherwise acceptable spectra solely due to poor phase estimates from sparse photometry. In many cases this involved fitting supplementary photometry obtained from other surveys. A summary of this investigation and the updated phase estimates can be found in Appendix~\ref{section:supplementary_photometry}. Following this investigation we are left with \LCFlagSpectra spectra for which we have reliable phase estimates. As we are specifically interested in the evolution of \SiFeature\ high-velocity components, we limit our search to the pre-peak regime when these features are expected to be present. Cutting any spectra with a phase post-peak, we are left with \PrepeakSpectra spectra from \PrepeakSNe SNe Ia.}

\begin{figure}
    \includegraphics[width = \columnwidth]{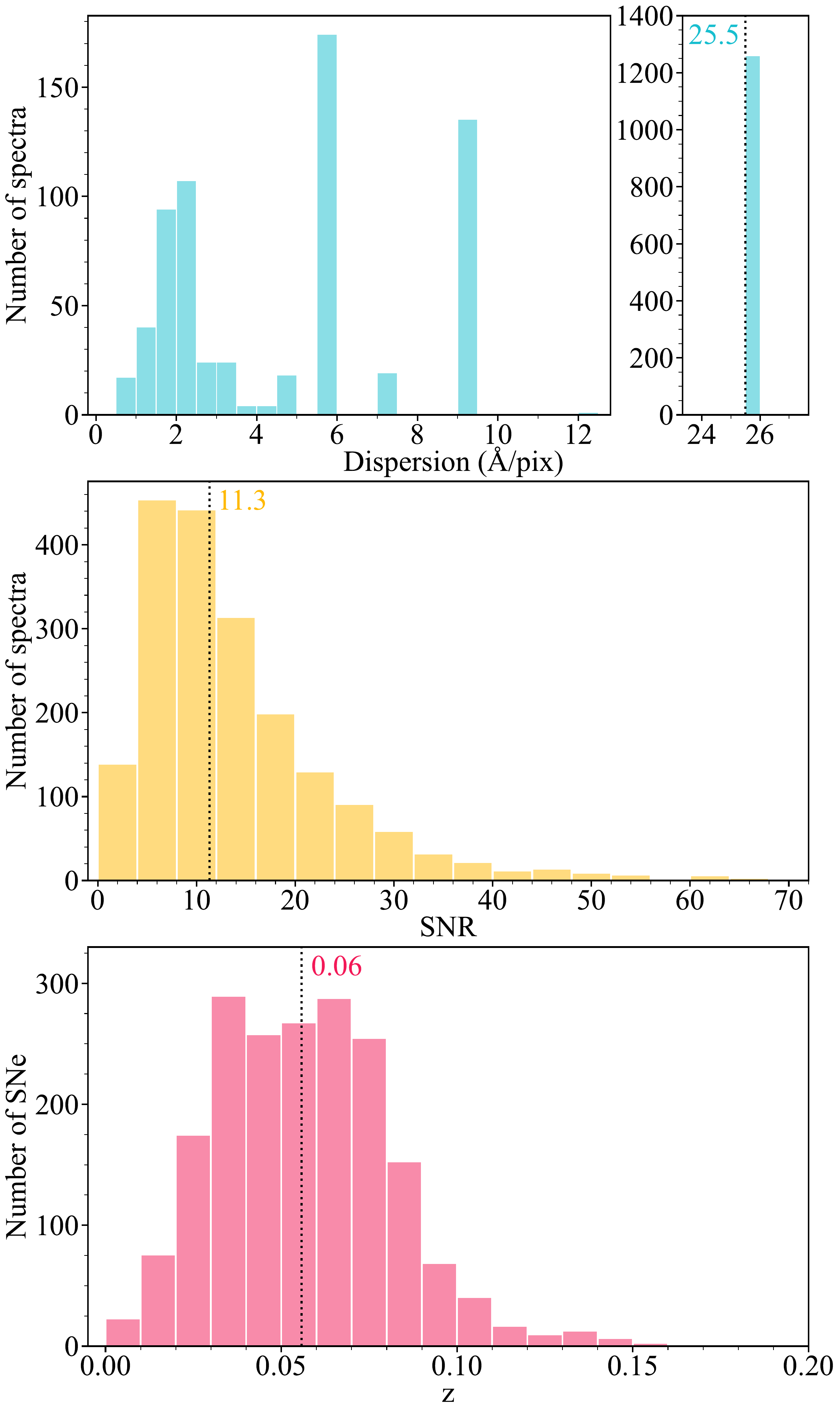}
\caption{The distributions of resolution and SNR in the \VisibleSiSpectra spectra before peak, covering the relevant wavelength region with a clean enough signal for the SNR estimation. The bottom panel represents the redshift distribution of the \VisibleSiSNe objects sampled by these spectra. The dotted lines and accompanying values correspond to the medians of the three measurements. Final cuts upon SNR and resolution will be implemented based upon the results of the simulated fits.}
    \label{fig:snr_resolution_hists}
\end{figure}

To assess the spectral quality in the vicinity of the \SiFeature\, we define a local signal-to-noise ratio (SNR) as the ratio of the depth of the \SiFeature\ to the standard deviation in the regions of the continuum to the red and blue (continuum region selection shall be discussed in detail in Sect.~\ref{sec:method:hvf_search}). This measurement is made once the continuum has been removed and is repeated 1000 times with different continuum selection points. The final SNR estimate is taken as the mean of these 1000 values. In making these estimates we identified \FailedWlCoverageInvisibleSiSpectra\ spectra that either did not cover the wavelength range of the \SiFeature\ feature, or did not have a clearly visible \SiFeature\ feature from which to compute an estimate of local SNR. These spectra were cut from the sample and the measurements of SNR for the remaining objects can be found in Fig.~\ref{fig:snr_resolution_hists}, along with the distribution of \textcolor{black}{dispersion (average separation between pixels in \AA)} and the redshift distribution of the corresponding SNe. The large peak at $\sim$25~\r A in the spectral \textcolor{black}{dispersion} panel corresponds to the spectra coming from the Spectral Energy Distribution Machine \citep[SEDM;][]{SEDM, SEDM_Mickael} on the P60 \citep{Cenko2002_P60} at the Palomar Observatory. The SEDM provides $\sim$60 per cent of the full DR2 spectral sample. The usefulness of the SEDM spectra in the search for HVFs will be determined by the simulations in Sect.~\ref{sec:method:simulations}, in which we will place a cut on both \textcolor{black}{dispersion} and SNR.


\section{Method}
\label{sec:method}

In Section.~\ref{sec:method:hvf_search}, we detail the procedure for fitting the \SiFeature\ and our method of classifying high-velocity and non-high-velocity features. In Section \ref{sec:method:simulations}, we describe the formulation of our synthetic \SiFeature\ features, as well as present the simulation results that test the ability of our method to identify HVFs. Based on these results, we can determine the \textcolor{black}{dispersion} and SNR that are required to identify HV components in observed spectra. We also investigate the simulation results to assess potential biases in our measurements of the feature parameters.

\subsection{Identification of high-velocity components in the \SiFeature}
\label{sec:method:hvf_search}
Our aim is to distinguish between those SN Ia spectra that possess a HV component in the \SiFeature\ line and those that do not. The feature fitting with single and double component models shall be performed with an Markov-Chain Monte Carlo (MCMC) framework using the \textsc{emcee} package \citep{emcee}, treating the local continuum as linear and leaving the corresponding slope and intercept as free parameters in the fitting.

\subsubsection{Spectral line fitting}
\label{sec:method_spectral_line_fitting}
As for the SNR estimation discussed in Section \ref{sec:sample_def}, the line fitting requires defined continuum regions to the blue and to the red of the \SiFeature\ absorption feature. The so-called `continuum' in these photospheric phase spectra is in fact the overlapping of many different P-Cygni emission profiles and as such we are actually defining a pseudo-continuum. This pseudo-continuum region selection is more complex for the observed data than for the simulated features for reasons such as contamination by neighbouring features and noise spikes. The initial selections of these local pseudo-continuum regions were performed using a gradient method similar to previous studies (e.g. \citealt{blondin2011, nordin2011, berkeley, silvermanHVF}). Starting at the local minimum of the feature, gradients are calculated in wavelength bins successively moving outwards to the red and blue until the gradient changes sign, indicating a maximum. The wavelength bins corresponding to these sign changes are therefore taken as the initial pseudo-continuum regions selections. Each selection was subsequently checked over manually, with any necessary updates to the regions being made. 

For each spectrum, we commence with a pre-processing step which involves the cutting of two small regions of the spectrum (6275--6307~\r A and 6860--6890~\r A) corresponding to telluric regions, as well as correcting for the host galaxy redshift. We then perform a normalisation step by dividing the flux of the spectrum by the maximum value found between 200~\AA\ to the blue of the blue continuum region and 200~\AA\ to the red of the red continuum region. This ensures that the slopes and offsets of all the features in the fitting will be of a similar order of magnitude.

The \SiFeature\ feature is a doublet comprised of two lines very close together in wavelength space at 6347.11 and 6371.37~\AA. Our feature fitting assumes the two singlets of the \SiFeature\ doublet to be tied in all parameters (velocity, depth, and width) as is the case under the assumption of an optically thick regime at these pre-peak phases \citep[see discussion in][]{childress2013}. The single component model corresponds to one \SiFeature\ doublet, typically associated with the photosphere and therefore, we call this the photospheric-velocity (PV) component. The double-component model includes an additional \SiFeature\ doublet at higher velocity, which we refer to as the high-velocity (HV) component.

The single ($f_1$) and double ($f_2$) component models take the forms,
\begin{equation}
\centering
    f_1 = (1-g(\lambda, a_{\text{PV}}, b_{\text{PV}}, c_{\text{PV}}))(s \lambda + i)
\label{eqn:single_component_model}
\end{equation}
and
\begin{equation}
\centering
    f_2 = (1-g(\lambda, a_{\text{PV}}, b_{\text{PV}}, c_{\text{PV}})-g(\lambda, a_{\text{HV}}, b_{\text{HV}}, c_{\text{HV}}))(s \lambda + i)
\label{eqn:double_component_model}
\end{equation}
with $f$ as the flux, $s$ and $i$ as the continuum slope and intercept respectively, and $g(\lambda, a, b, c)$ being an indivdual \SiFeature\ doublet of the form:
\begin{equation}
\centering
    a\exp{\frac{-(\lambda-(b-7.89))^2}{2c^2}}+a\exp{\frac{-(\lambda-(b+16.37))^2}{2c^2}}.
\label{eqn:si_doublet_equation}
\end{equation}
where $a$ is the depth of one of the singlets, $b$ is the wavelength position of the minimum, with the 7.89 and 16.37 quantities as the offsets from 6355~\r A for the 6347.11~\r A and 6371.37~\r A lines of the doublet, $c$ is the width, and the PV and HV subscripts refer to the photospheric- and high-velocity components of the model, respectively. The uncertainty on the flux values is taken as the standard deviation of the points in the continuum regions around the linear fit used for the initialisation of $s$ and $i$.

We place a number of priors on the MCMC fitting to restrict the walkers to the relevant region of the parameter space. The depth parameters ($a_\text{PV}$ and $a_\text{HV}$) and width parameters ($c_\text{PV}$ and $c_\text{HV}$) are bound to be larger than 0.05 and 30~\AA, respectively to avoid overfitting, especially in the noisier and lower resolution spectra. To ensure that the fitted components fall within the wavelength region of the feature we introduced the restrictions that the sum $b_\text{PV}+c_\text{PV}$ was smaller than the lower bound of the red continuum region and than $b_\text{HV}-c_\text{HV}$ was higher than the upper bound of the blue continuum region (or $b_\text{PV}-c_\text{PV}$ in the single component model).

Through preliminary testing with simulated features (see Section \ref{sec:method:simulated_data}), we found that in most cases the slope from a linear fit to the continuum regions provided a smaller residual with the true value than that of the final output of the MCMC. This effect was counteracted by placing a uniform prior on the slope parameter, with \textcolor{black}{dispersion}-dependent limits $d$ away from this slope estimate derived empirically as
\begin{equation}
\centering
    d = (6\times10^{-6})r + (4\times10^{-5})
\label{eqn:slope_bounds}
\end{equation}
with $r$ as the \textcolor{black}{dispersion in \AA/pix}.

The measured values for the parameters are taken as the medians of the posteriors, with the lower and upper uncertainties as the 16th and 84th percentiles, corresponding to a 68 per cent confidence interval. We use the MCMC chains to calculate parameters such as pseudo-equivalent widths (pEWs), pEW ratios, and velocity separations directly, drawing the median and lower and upper uncertainties from the resulting posteriors. The accuracy of these uncertainties shall be investigated with the simulated data in Section~\ref{sec:method:simulations}.

\begin{figure}
    \includegraphics[width = \columnwidth]{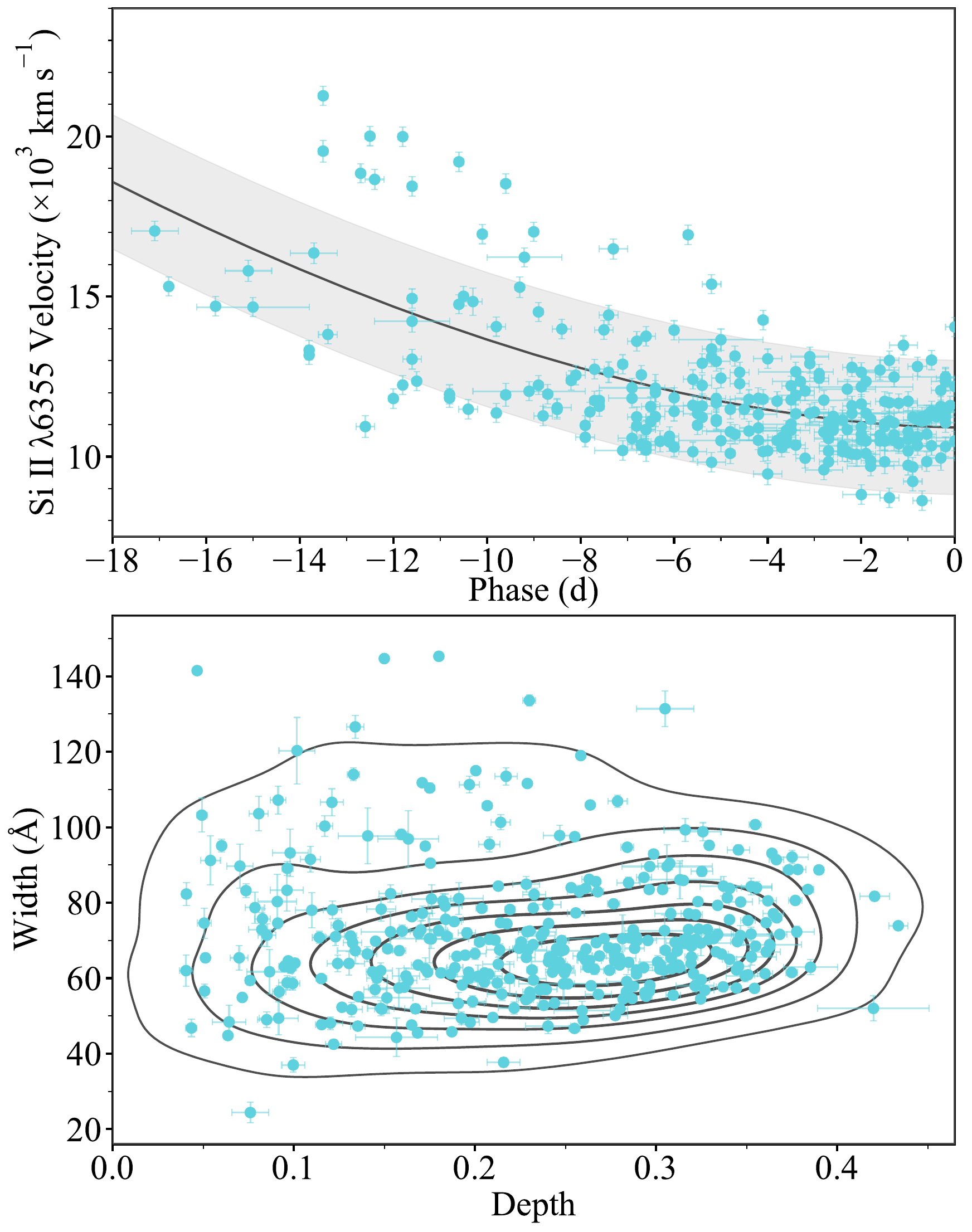}
    \caption{The PTF \citep{ptf} SN Ia spectral measurements for velocity, width and depth in the case of the \SiFeature\ absorption feature. The power law fit to the velocity evolution and the Gaussian KDE (described by the contours) for the width and depth are used to inform the generation of the synthetic features in the simulations.}
    \label{fig:ptf_data}
\end{figure}

\begin{figure*}
 \includegraphics[width = \linewidth]{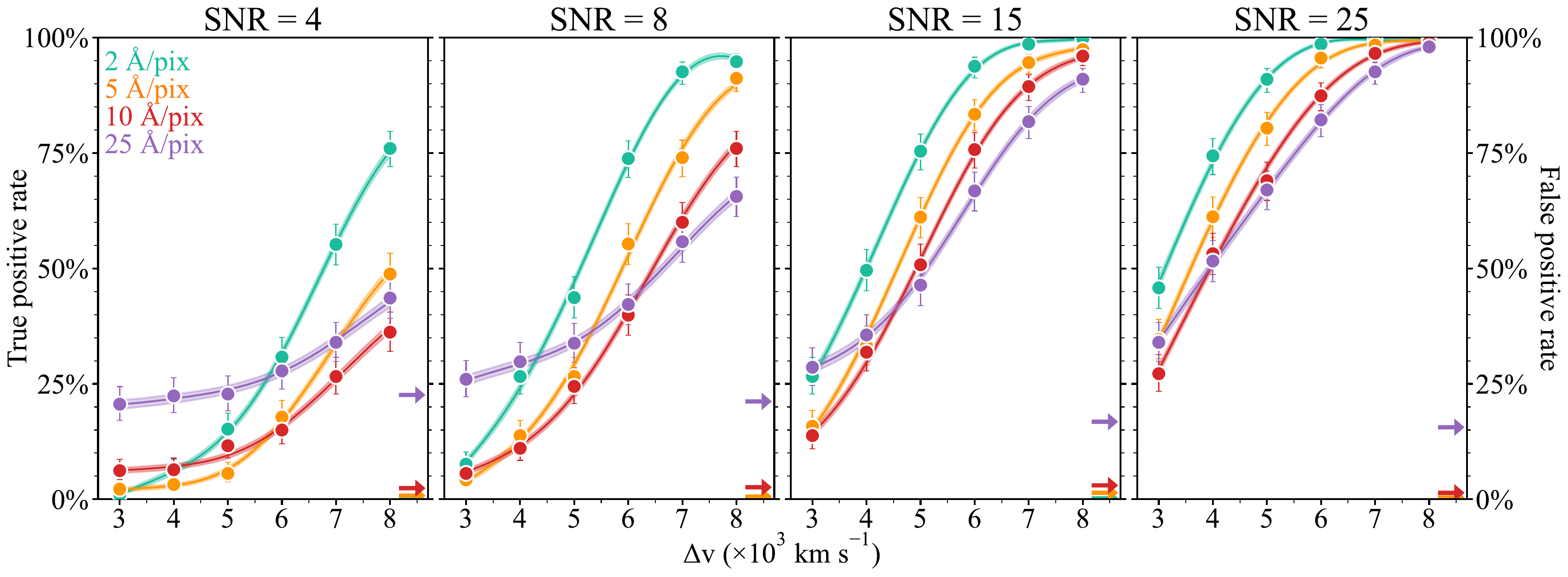}
 \caption{The true (circles) and false (arrows) positive rates of the MCMC/BIC classification method as a function of the velocity separations derived from the simulations for increasing SNRs from the left to the right panel and different spectral \textcolor{black}{dispersions} shown as different colours. 1D slices (in $\Delta v$ space) of the 3D GP interpolation of SNR, \textcolor{black}{dispersion}, and velocity separation are presented as the coloured lines and associated 68\%\ confidence interval regions.}
 \label{fig:true_false_positive_rates}
\end{figure*}

\subsubsection{Deciding upon the preferred model}
\label{sec:method:deciding_upon_preferred_model}
With single- and double-component posterior distributions derived from the MCMC, we now employ the Bayesian Information Criterion (BIC) - a goodness-of-fit metric that disfavours more complex models - to decide upon the preferred model. As there are three more free parameters in the double-component model than the single-component model, the corresponding fit has to significantly improve the goodness-of-fit in order to be preferred by the BIC. We employ the BIC as
\begin{equation}
\centering
    \text{BIC} = -2\ln(L) + k\ln(N)
\label{eqn:bic_formulation}
\end{equation}
with $L$ as the maximised value of the likelihood function, $k$ as the number of free parameters (5 for the single and 8 for the double component), and $N$ as the number of data points in the fitting. The model that produces the lowest BIC is provisionally deemed to be the best match, subject to two cuts explained below.

HV components that possess large velocity separations from their PV counterparts are easily distinguishable through fitting or visual inspection. However, as this velocity separation decreases, the components become increasingly more entangled and more difficult to detect. With the addition of noise, this has the potential to lead to high false-positive rates. Therefore, a cut upon the minimum velocity separation is necessary in the fitting, as below this limit we will not be able to reliably differentiate between single and double component features. A lower limit of 4500~\kms\ was adopted in \cite{silvermanHVF}. Our lower limit of 4000~\kms\ was set based on the performance of our classification method on simulated data (Section \ref{sec:method:simulations}). Similarly to \cite{silvermanHVF}, we consider any two-component classifications with a PVF velocity $\leq 9000$~\kms\ as unreliable, with the fitted PV component likely corresponding to contaminating \CII\ absorption. Therefore, in these cases we choose the single- over double-component classification, regardless of the BIC-based classification.


\subsection{Simulations}
\label{sec:method:simulations}

To investigate the accuracy of the MCMC/BIC method described in Section \ref{sec:method:hvf_search} for identifying HV components, as well as its ability to recover the injected parameters, we generated synthetic \SiFeature\ features. The results from these simulations were also used to inform cuts on \textcolor{black}{dispersion} and SNR in the observed data. In Section \ref{sec:method:simulated_data}, we describe the construction of our simulated data for testing our method, while in Section \ref{sec:method:detection_efficiencies}, we detail our detection efficiencies and in Section \ref{sec:method:parameter_recovery}, we discuss the correction for biases in our fitting based on our simulations.

\subsubsection{Constructing the simulated data}
\label{sec:method:simulated_data}
The synthetic features are comprised of two (PV and HV) doublet components, separated by a fixed velocity difference - with the exception of the simulations where we injected no HV component to assess the false-positive rate. We created a grid spanning four SNRs (4, 8, 15, and 25), four \textcolor{black}{dispersions} (2, 5, 10, and 25~\AA/pix) to cover the range of values found for the DR2 subsample (Fig.~\ref{fig:snr_resolution_hists}), as well as seven velocity separations (no HVF, 3000, 4000, 5000, 6000, 7000, and 8000~\kms) covering the range of separations expected from the data. For each of these parameter setups we generated 500 simulated features.

Measurements from PTF SNe Ia \citep{ptf} were used to inform our generation of \SiFeature\ line profiles. For each feature we generate a random phase between $-$17 and 0~d, which then gives a photospheric velocity probability distribution as the cross-section of the power-law fit to the PTF data seen in the top panel of Fig.~\ref{fig:ptf_data}. We draw a photospheric velocity from this resulting distribution, with the HVF velocity then defined by the fixed velocity separation in the simulation. The depths and widths of both components are drawn from a kernel density estimate (KDE) of the PTF measurements found in the bottom panel of Fig.~\ref{fig:ptf_data}. The modelling of the HV components in these simulations is rather simplistic, with the depths and widths assumed to follow the same distribution as those of the PV components. The photospheric velocity, width, and depth estimates based on \cite{ptf} are also biased since these include contributions from high-velocity features.  The validity and potential implications of these assumptions can be assessed with our final DR2 sample and shall be discussed in Section~\ref{sec:results:phase_evolution}.

Noise was introduced to the simulated features and surrounding continuum based on the set SNR values defined by the simulation grid. Noise values for each wavelength bin were then drawn from a Gaussian distribution centered about zero with a standard deviation as the product of the SNR and the depth of the composite feature. The \textcolor{black}{dispersion} of each simulated feature is simply defined as the separation between the wavelength bins (the same definition as used for the observed spectra). With this simulated feature constructed, we introduced a simple linear continuum to each feature, varying the slope and intercept between them. The values for these slopes were drawn from the normal distribution with mean $-1.9\times10^{-4}$ and standard deviation $1.7\times10^{-4}$, chosen to match the range of initial slope estimates from the normalised DR2 spectra.

Our simulated spectra with the single- and double-component models were then fit following the procedure described in Section~\ref{sec:method:hvf_search}. The continuum region selection for the simulated data was performed in a similar fashion to that of the observed data but did not require manual checking for cosmic ray spikes or contamination by host lines. 

\subsubsection{Detection efficiencies}
\label{sec:method:detection_efficiencies}

\begin{table}
\caption{Breakdown of the number of spectra/SNe Ia remaining after each cut to reach the final sample of \FinalSpectra\ spectra ready for fitting.}
\label{tab:cuts}
\begin{tabular}{lcc}
\hline
\textbf{Cut} & \textbf{Spectra} & \textbf{SNe} \\ \hline 
\textit{Initial cuts} & & \\ \hline
ZTF Cosmo DR2 & \FullSampleSpectra & \FullSampleSNe \\
Light curve quality cuts & \LCFlagSpectra & \LCFlagSNe \\
Phase \textless~0~d & \PrepeakSpectra & \PrepeakSNe \\
\SiFeature\ wavelength coverage & \WlCoverageSpectra & \WlCoverageSNe \\
Visible \SiFeature\ & \VisibleSiSpectra & \VisibleSiSNe \\ \hline
\textit{Spectral quality cuts} & & \\ \hline
\textcolor{black}{Dispersion $\leq$~10~\AA/pix} & \ResolutionCutSpectra & \ResolutionCutSNe \\
SNR $\geq$~8 & \SNRCutSpectra & \SNRCutSNe\\ \hline
\end{tabular}
\end{table}

We have performed simulations of the \SiFeature\ region with different SNRs, \textcolor{black}{dispersions} and the presence (at different velocity separations between the low- and high-velocity components) or absence of a HV component. We present the resulting true- and false-positive rates for the detection of high-velocity features for each of the different simulation setups in Fig.~\ref{fig:true_false_positive_rates}. The panels display the four SNR ratios increasing from left to right, with the different colours indicating the four different \textcolor{black}{dispersions}. The circle markers correspond to the left axis and display the percentage of the 500 simulated features in that simulation that were correctly identified as possessing both a high-velocity and photospheric velocity component. The arrows correspond to the right axis and represent the percentage of the 500 simulated features without a HV component that were incorrectly identified as possessing two components. The uncertainties on these points are taken as binomial and calculated from the Clopper-Pearson method with a 95\% confidence interval.

Each simulation setup in our grid can be thought of as a point in a 3D parameter space spanning SNR, \textcolor{black}{dispersion}, and velocity separation. At every point in this space there exists some true-positive rate which can be used to assess how probable it is to correctly identify a HV component with such properties. Using the Gaussian Process (GP) module within \textsc{scikit-learn} \citep{sklearn} we perform a 3D interpolation of this discrete grid to obtain a continuous view of the changing detection efficiency. One-dimensional slices in velocity space are shown in Fig.~\ref{fig:true_false_positive_rates}.

\begin{figure}
 \includegraphics[width = \linewidth]{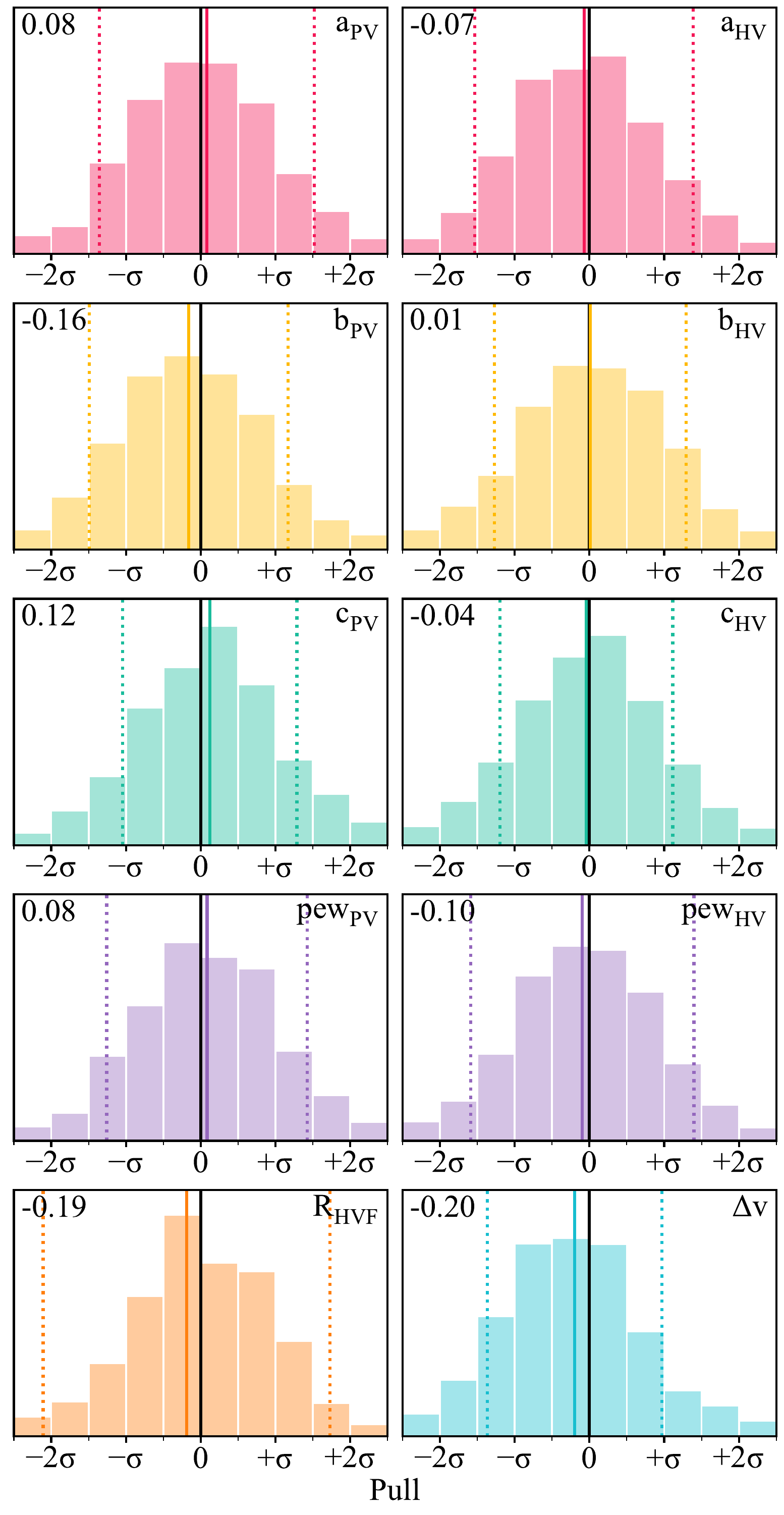}
 \caption{The distribution of the pulls (residual/uncertainty) for the six feature parameters in the case of spectra with a SNR of 15, a \textcolor{black}{dispersion} of 5~\AA/pix and a velocity separation of 5000~\kms. The solid black lines indicate the desired zero pull value, while the solid coloured lines indicate the means of the distributions, the values of which are shown in the top left corner of each panel. The dotted lines display the measured standard deviations.}
 \label{fig:pull_distributions}
\end{figure}

\begin{figure*}
 \includegraphics[width = \linewidth]{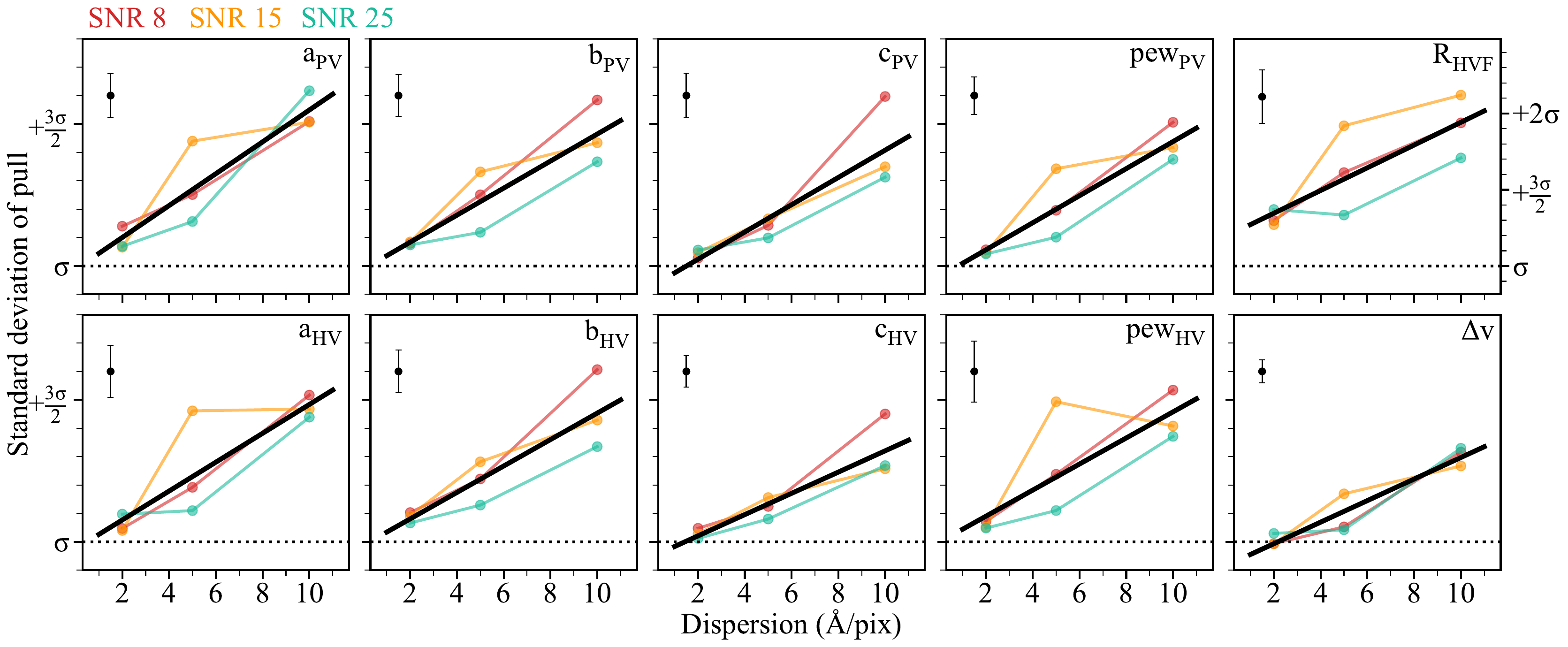}
 \caption{The standard deviations of the pull distributions about their means for each of the simulations. The three SNRs are represented by different colours as indicated above the top left panel. The thick solid lines describe the linear regression fits to the datapoints with each of the three \textcolor{black}{dispersions}. The black point in each panel presents the standard deviation of the points around this fit. We highlight that the R$_\text{HVF}$ panel has a different y-axis scaling due to the larger standard deviation values.}
 \label{fig:uncertainty_corrections}
\end{figure*}

Our aim is then to introduce cuts based on the velocity separation, SNR, and \textcolor{black}{dispersion} to keep the true-positive rate high and the false-positive rate to a minimum, while maximising our sample size.  As one might expect, we see an increase in our ability to correctly identify HVFs as the noise is reduced, as we improve the spectral resolution \textcolor{black}{(decrease the dispersion)}, and as we separate the components in velocity space. The \textcolor{black}{highest dispersion} simulations (purple in Fig.~\ref{fig:true_false_positive_rates}) have a separation of 25~\AA\ between wavelength bins and represent the large percentage of DR2 spectra coming from the SEDm spectrograph. Accompanying the smaller true-positive rates for these features, we see large false-positive rates of the order 15~-~20\%. These false-positive rates also stand as lower thresholds for the true-positive rates as we can see for the 25~\AA/pix lines in the left hand panel, where the true-positive rates flatten off and plateau at the level of the false-positive rate as we approach smaller velocity separations. This leveling off of the true-positive rate as it becomes similar in magnitude to the false-positive rate causes an overlap effect between the 25~\AA/pix simulations and those with \textcolor{black}{smaller dispersions}. The simulations with SNR of 4 exhibit the same overlap effect as seen for the 25~\AA/pix \textcolor{black}{dispersion} simulations for the 10 \AA/pix \textcolor{black}{dispersion} spectra for velocity separations up to 5000~\kms.

Due to these issues with the lowest resolution and lowest SNR spectra, we cut all spectra from our observed sample with a \textcolor{black}{dispersion} greater than 10~\AA/pix and SNR~$<$~8. With this \textcolor{black}{dispersion} cut, we remove all SEDm spectra from our sample, which due to its brighter limiting magnitude for targets compared to the other telescopes, may potentially remove brighter events from the sample. We test the impact of this on our demographic comparison in Section \ref{sec:results:hvf_properties} but find no bias in SALT2 $x_1$ for our sample compared to the volume-limited DR2 sample. Our SNR cut could also potentially bias our sample, removing events with intrinsically shallower \SiFeature\ features. However, `shallow Silicon' \citep{branch2006} SNe Ia have light curve widths (parameterised by e.g., SALT2 $x_1$) that are higher than mean of the SNe Ia sample \citep[e.g.][]{blondin2011} and we find no $x_1$ bias relative to our comparison DR2 sample, again suggesting that this SNR cut does not introduce a significant bias to our final sample.  With these two spectral quality cuts in place, we obtain our final sample of \FinalSpectra\ observed spectra from \FinalSNe\ SNe ready for fitting. The breakdown of each of the cuts can be found in Table~\ref{tab:cuts}.

The final result to be drawn from Fig.~\ref{fig:true_false_positive_rates} is a cut upon the measured velocity separation based on the simulations. \cite{silvermanHVF} chose to disregard any classifications of features with separations less than 4500~\kms, \textcolor{black}{instead classifying these spectra as having single components}. We chose a similar threshold of 4000~\kms, as below this velocity we begin to see the previously discussed overlap effect below 4000~\kms\ for the 10~\AA/pix simulations  with SNR of 8 (lowest final resolution and SNR studied). \textcolor{black}{This is further justified by closer inspection of the false positive rate. Averaging over all simulations passing the implemented dispersion and SNR cuts, we find a false positive rate of $\sim$1\%, with 80\% of these false positive classifications measured with velocity separations less than 4000~\kms. Applying this velocity separation cut reduces the simulation false positive rate to $\sim$0.2\% in the remaining simulations. Given that the simulations represent the ideal scenario with linear continua, generated Gaussian features, and Gaussian noise, this rate is optimistic; therefore, we assume a more conservative false positive rate of 2\%. The false positives as measured in the DR2 are likely to be concentrated at smaller velocity separations up to $\sim$5000~\kms\ where there exists more degeneracy between the single and double component models.}

\subsubsection{Parameter recovery and uncertainty corrections}
\label{sec:method:parameter_recovery}

Along with estimating the detection efficiencies of our MCMC/BIC HVF classification method, the grid of simulations also allows us to test how the estimated uncertainties perform in terms of truly representing 68\% confidence intervals, and identify potential biases in our measurements. The pull for some parameter $X$ is calculated as
\begin{equation}
    \text{Pull}_X = \frac{{X}_{\text{fit}}-{X}_{\text{simulated}}}{\sigma_{X}}
\end{equation}
with $\sigma_X$ as the uncertainty taken from the posterior distribution. In the case of no biases and uncertainties that truly reflect 68\% confidence intervals, these distributions should be Gaussian with a mean of zero and a standard deviation of unity.

\begin{figure*}
 \includegraphics[width = \linewidth]{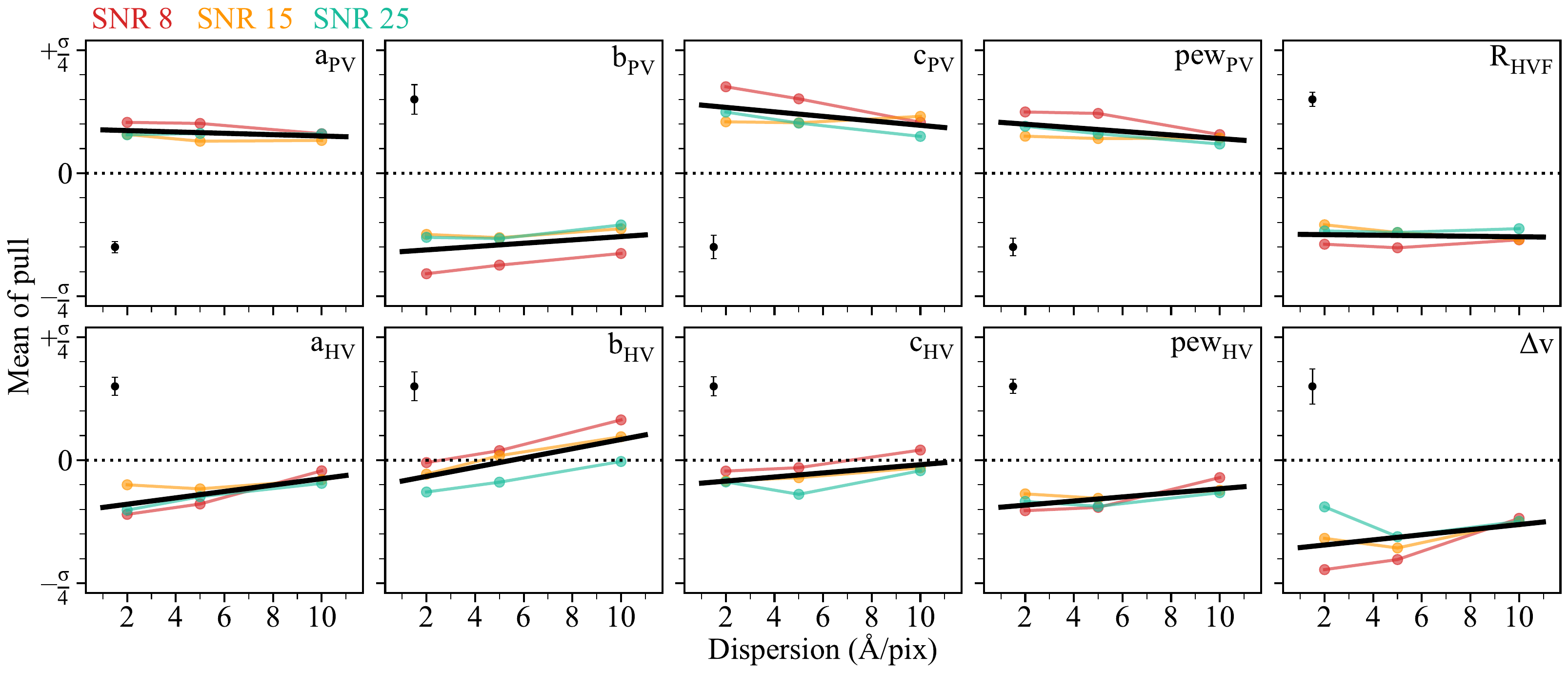}
 \caption{The means of the pull distributions for each of the simulations after applying the uncertainty corrections. The formatting matches that of Fig~\ref{fig:uncertainty_corrections}. The thick black lines present linear regressions to these trends which can be used to characterise and then correct for the biases. The black point in each panel presents the standard deviation of the points around this fit.}
 \label{fig:pull_corrections}
\end{figure*}

In Fig.~\ref{fig:pull_distributions} we show example distributions of the pulls in the six parameters describing the PV and HV components, as well as the pEWs, R$_\text{HVF}$ (the ratio of pEW$_\text{HV}$/pEW$_\text{PV}$), and $\Delta v$ from a simulation with our middle remaining \textcolor{black}{dispersion} of 5~\AA/pix and middle remaining SNR of 15, with a velocity separation of 5000~\kms. As clearly visible for the majority of the measured parameters, the spread of the pull distribution is too large with a mean standard deviation of $\sim1.3$ about the mean for this set of simulation parameters, indicating that the uncertainties from the MCMC posterior are $\sim1.3$ times too small. A pull is also observed in most parameters where the means of the distributions do not lie exactly at zero.

In Fig.~\ref{fig:uncertainty_corrections}, we present the standard deviations for each of the parameters as a function of \textcolor{black}{dispersion} and split into the three remaining SNR (8, 15 and 25). All two-component fits with resulting velocity separation measurements below the 4000~\kms\ threshold have been already removed. Most of the 2~\AA/pix \textcolor{black}{dispersion} simulations exhibit standard deviations close to unity and therefore, require little to no correction. The standard deviations of the pull generally increase with \textcolor{black}{increasing dispersion} for all SNR simulations. The thick black lines present linear regressions to these combined SNR standard deviation trends and are used to compute scale factor corrections to the uncertainties to bring them in line with 68\%\ confidence intervals.

In a similar fashion, we can analyse how closely the means of these pull distributions lie to zero to identify any biases in the measurements. In Fig.~\ref{fig:pull_corrections} we plot the means of the pull distributions for each of the simulations as a function of \textcolor{black}{dispersion}, after having applied the uncertainty corrections discussed above. We observe slight biases that are consistent between the different SNR and \textcolor{black}{dispersion} pairings and appear relatively flat with changing \textcolor{black}{dispersion}. We correct for these trends using a linear regression as a function of \textcolor{black}{dispersion}. The same investigation was performed for the single component fits to the simulated features with no HV components. While we find the means to be centered upon zero -- indicating no bias in the measurements -- the standard deviations of these distributions fell in the range 1 -- 1.3 and therefore, small corrections were made to the uncertainties using the same method as for the two-component fits above.

\begin{figure}
 \includegraphics[width = \linewidth]{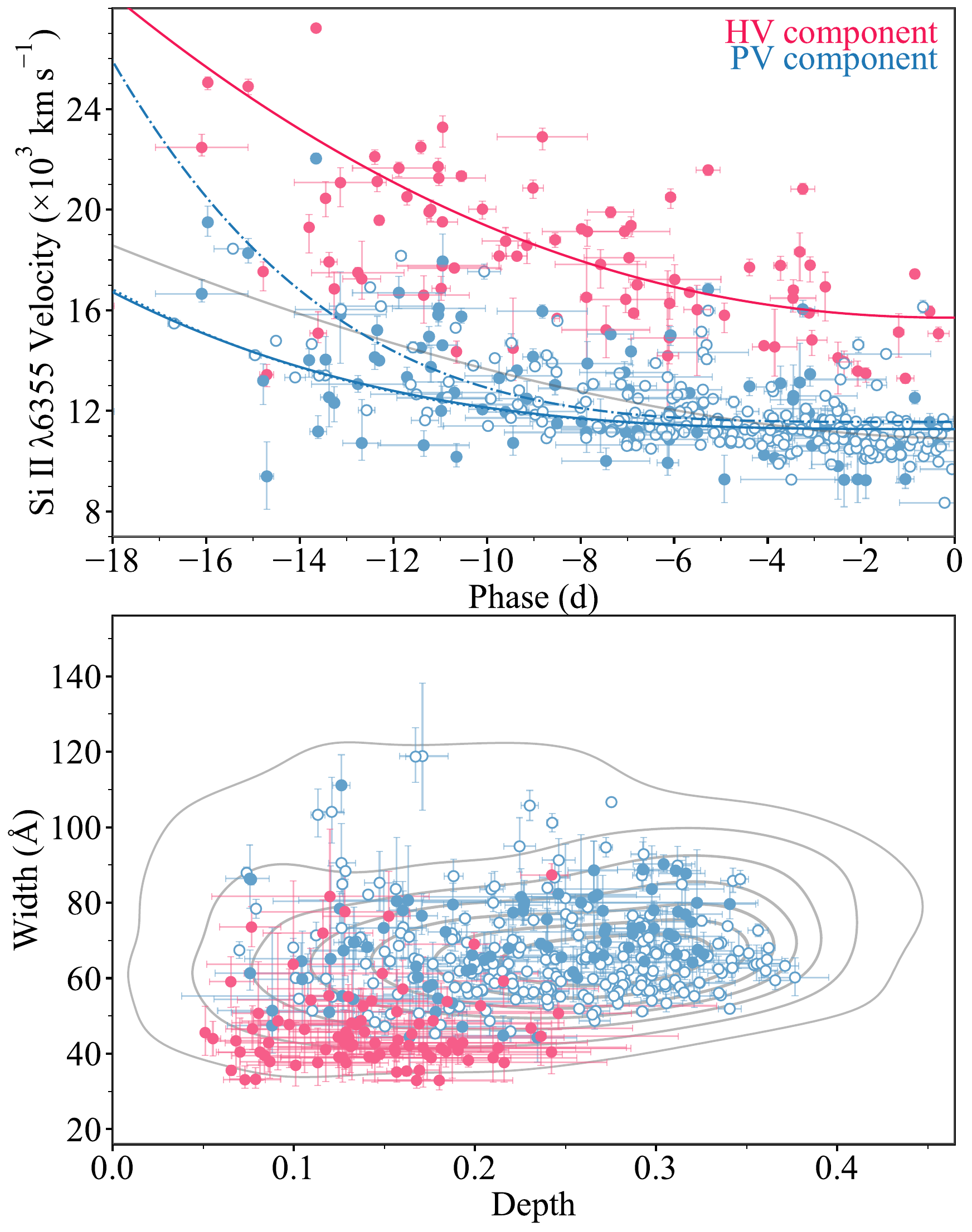}
 \caption{The evolution of the velocities of the PV and HV \SiFeature\ components for all \FinalSpectra\ spectra in the sample (top panel). \textcolor{black}{The hollow points correspond to the \PVFSpectra\ PV components with no HV counterpart.} The solid blue line is fit to all the PV components, with the dash-dotted and dotted blue lines showing fits to the PV components of spectra with and without HV components, respectively. The pink line is the fit to the HV components and the grey line is the evolution from the PTF sample \protect{\citep{ptf}}. The width against depth for the measured PV and HV \SiFeature\ components using the same colours as the top panel (bottom panel). The grey contours show the KDE for the PTF sample for comparison. }
 \label{fig:velocity_evolution}
\end{figure}


\section{Results}
\label{sec:results}

With the cuts informed by the simulations (\textcolor{black}{dispersion~$\leq$10~\AA/pix} and SNR~$\geq$8), we arrive at our final sample of \FinalSpectra\ spectra (see Table~\ref{tab:cuts}) for which we can study the presence of high-velocity features in the \SiFeature. With the MCMC/BIC classification method outlined in Section~\ref{sec:method:hvf_search}, we fit our sample, taking all those classified as HVF with $v_\text{PV}~>~9000$~\kms\ and $\Delta v~>~4000$~\kms\ as our HVF subsample.
Of the \FinalSpectra\ spectra we identify \HVFSpectra\ as possessing both PV and HV components, corresponding to \HVFPercentage\% of the sample. The HVF spectral sample comprises spectra from \HVFSNe\ SNe, including 8 objects for which we find these HV components in multiple epochs. In the sample there exists one object (ZTF18abauprj) for which we have both HVF spectra, and a final epoch showing no sign of a HV component, providing a glimpse at the full evolution of these features as time evolves. In our sample, we also have 10 pairs of spectra from the same object coming from different instruments with phase separations of a day or less. In all these cases the classifications are consistent between these spectra with 6 pairs as PVF and 4 as HVF. The measured values for velocity for all these components are consistent within the uncertainties, as are the feature widths and the majority of the feature depths. Any differences in the depth parameters - which are all $\leq$0.05 in size - are likely predominantly the result of changing line profiles over the hours elapsed between the spectra, although there may be small systematic differences.

In Section \ref{sec:results:validation_priors}, we describe how the observed SN measurements were used to validate the priors used in the simulations. In Section~\ref{sec:results:phase_evolution} we explore the phase evolution of the HVFs. We then define and analyse a reduced low-bias sample to draw conclusions about the properties of HVFs in Section~\ref{sec:results:hvf_properties}, and correlations with light curve and host parameters in Section~\ref{sec:results:observables}.


\subsection{Validation of simulation priors}
\label{sec:results:validation_priors}

The top panel of Fig.~\ref{fig:velocity_evolution} displays the measured velocity evolution for the \FinalSpectra\ spectra from our \FinalSNe\ SNe Ia, split into the photospheric components, and high-velocity components where classified. Power-law fits were performed to the velocities as a function of phase for the HV components, as well as three samples of the PV velocities, i) all the PV components, ii) just the PV components with HV counterparts, and iii) just the singular PV components. These three fits are consistent from around $-$10~d onwards, with the PV velocities in objects possessing HV components drifting to higher velocities at early phases. This agrees with the findings of \cite{silvermanHVF} in that HVFs tend to be accompanied by higher velocity PVFs. We also plot in grey the velocity evolution from the PTF sample \citep{ptf} that was used as reference for our \SiFeature\ feature generation in the simulations. \textcolor{black}{These PTF measurements were performed on the overall \SiFeature\ features without distinction between PV and HV components. The PTF velocity evolution therefore likely has some level of contamination from HVFs, and would be expected to lie somewhere between our measured PVF (blue) and HVF (pink) evolution curves, as it does.}

\begin{figure}
 \includegraphics[width = \linewidth]{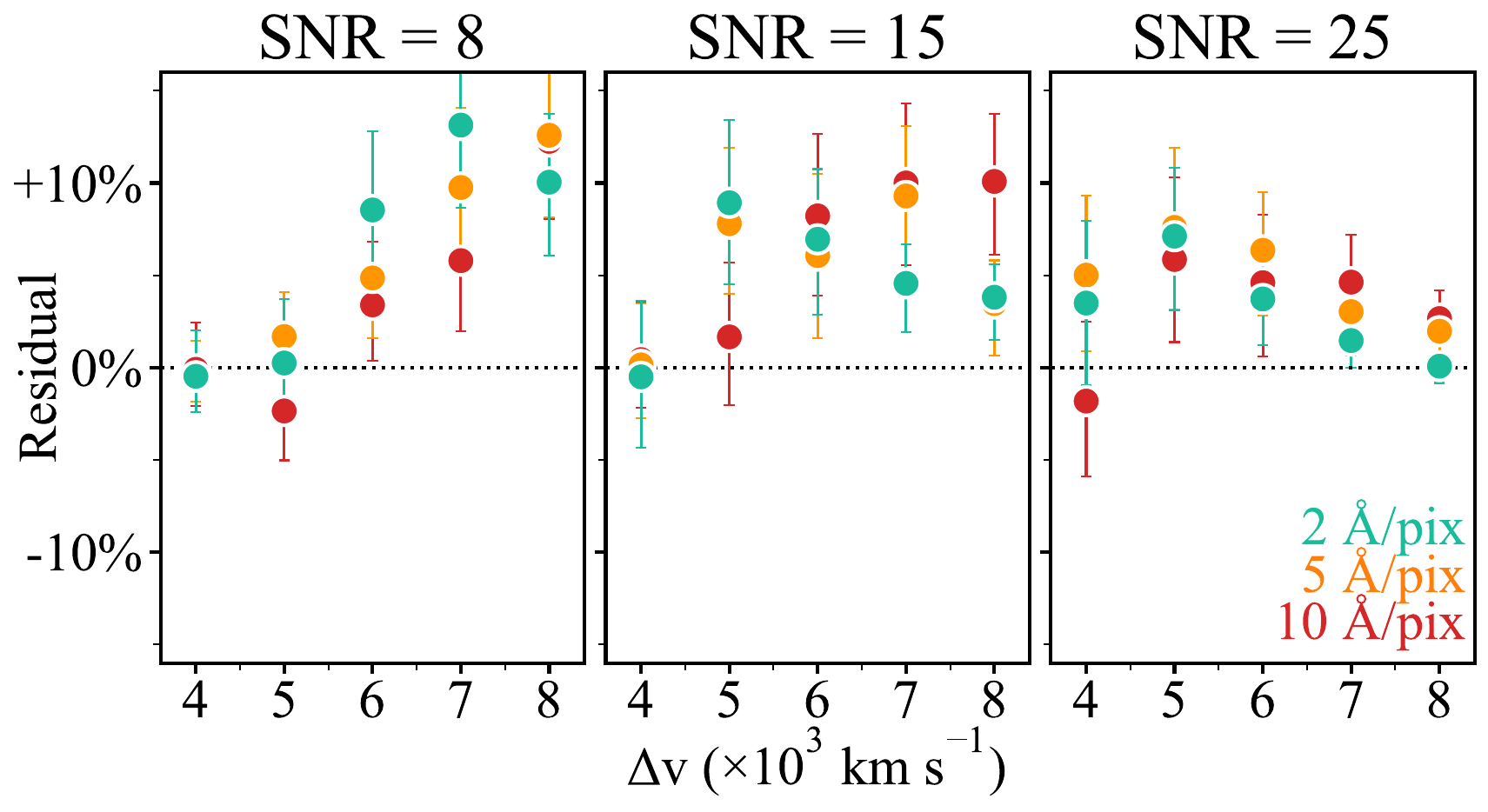}
 \caption{The residuals between the true positive rates taken from simulated spectra with $a_\text{HV}\leq 0.25$ or $c_\text{HV}\leq 70$~\AA, and the original true positive rates as a function of velocity separation for the three SNR, shown with increasing SNR from left to right panels.}
 \label{fig:simulation_cut_residuals}
\end{figure}

In the bottom panel of Fig.~\ref{fig:velocity_evolution} we present the measured depths and widths of the singlet lines making up the PV and HV components. As for the velocity evolution we plot the comparison PTF data as grey contours to show how our final measurements align with the distributions chosen to inform the feature generation in the simulations. As expected, the distribution of PVF measurements lie in the same parameter space as the PTF data, but our measurements are slightly more compact in both dimensions which, as with the velocity, is likely due to HVF contamination in the PTF dataset inflating the depths and widths in some cases. For the generation of the HVF components in the simulations, we made the assumption that the widths and depths of the HV components followed the same distribution. However, as clearly visible in Fig.~\ref{fig:velocity_evolution}, these components tend to cluster more towards shallower depths and narrower widths. \textcolor{black}{While some false positive classifications may be contaminating the HVF cluster, there remains a clear distinction between the HVF and PVF distributions.} To evaluate the effect that this discrepancy has upon the calculated detection efficiencies we recalculated the true-positive rates from the simulations, excluding any simulated spectra from the calculations with $a_\text{HV}>0.25$ or $c_\text{HV}>70$~\AA\ so has to be more in agreement with the measurements from the observations. The residuals between the true positive rates these updated simulations and the original simulations are presented in Figure~\ref{fig:simulation_cut_residuals}. In most cases, the true-positive rate increases with the cut to only focus on the observed region of the parameter space. This implies that we are less sensitive to finding HV components with similar widths and depths to their PV counterparts - likely due to increased degeneracy in the fit. Therefore, we updated the detection efficiencies with a GP interpolation (following the method of Section \ref{sec:method:detection_efficiencies}) to these new values.

\begin{figure}
 \includegraphics[width = \linewidth]{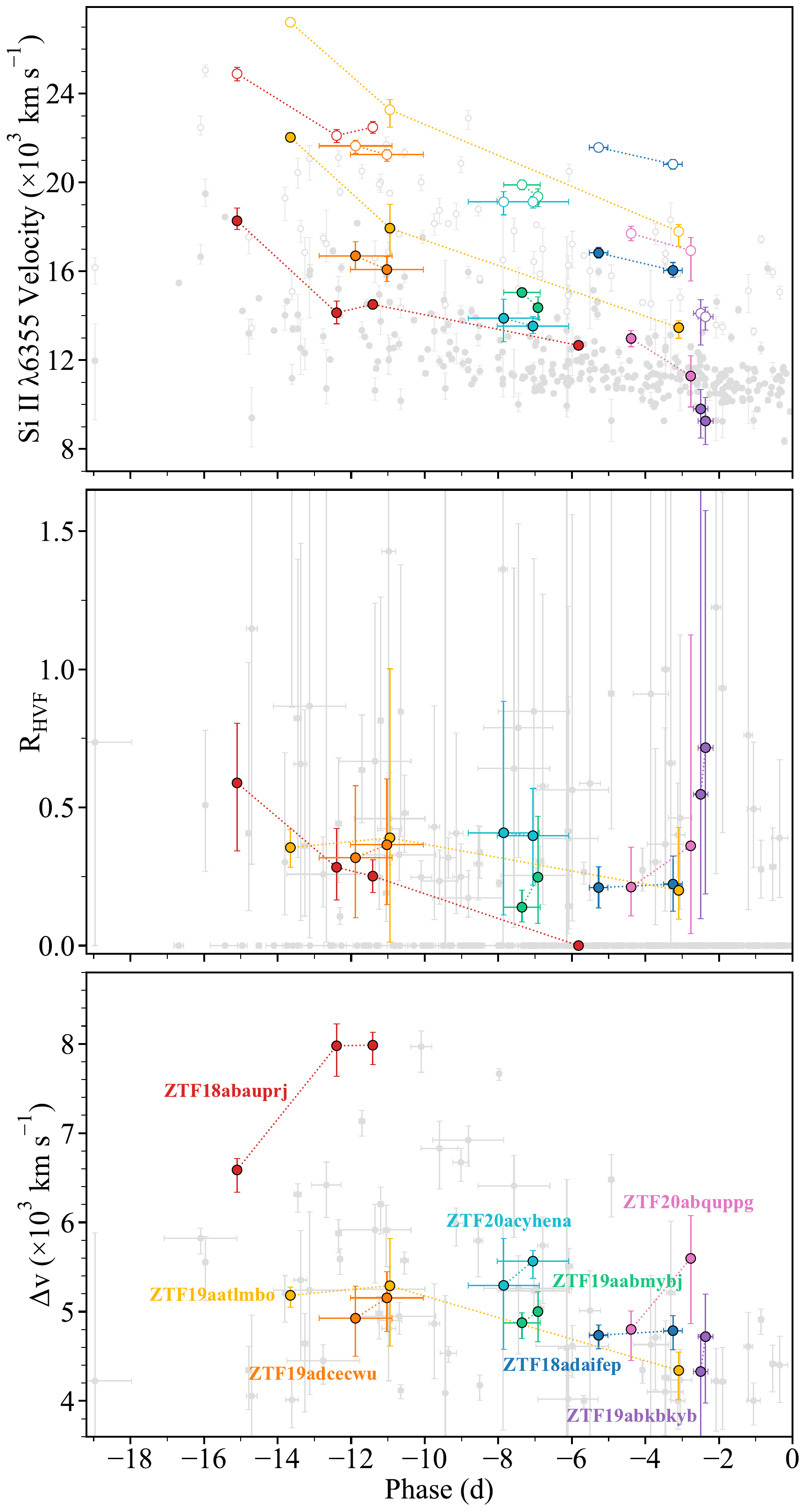}
 \caption{Evolution of the \textcolor{black}{component velocities (top)} parameter R$_\text{HVF}$ (pEW$_\text{HV}$/pEW$_\text{PV}$) (middle) and the separation of the PV and HV features in velocity space (bottom). The measurements for individual spectra are plotted in grey, with the coloured points and lines corresponding to SNe Ia for which we have multiple spectra. \textcolor{black}{The hollow points in the top panel correspond to the HV components.} The cluster of points in the middle panel at R=0 correspond to all the spectra identified as not having a HV component.}
 \label{fig:R_evolution}
\end{figure}

\subsection{Phase evolution}
\label{sec:results:phase_evolution}

The evolution of the ratio of pEW$_\text{HV}$/pEW$_\text{PV}$ (R$_\text{HVF}$) is seen for our observed sample in the \textcolor{black}{middle} panel of Fig.~\ref{fig:R_evolution}. In general, the uncertainties on the measured R$_\text{HVF}$ values are large as many of the two-component fits exhibit high levels of degeneracy and posteriors of the numerator and denominator in the ratio (pEW$_\text{HV}$ and pEW$_\text{PV}$) are inversely correlated.

\textcolor{black}{The SNe with multiple spectra allow us to probe the evolution of this ratio within individual objects, with the overall sample - including single-epoch objects - giving a more global view of how the distribution of R$_\text{HVF}$ changes with phase.} For most of our multi-phase objects, we possess only two spectral epochs, separated by less than $\sim$2.5~d, leading to very little evolution in R$_\text{HVF}$ which appears to remain approximately constant on such small timescales. However, for ZTF19aatlmbo (SN~2019ein) and ZTF18abauprj (SN~2018cnw),  we have three and four spectra spanning 9.3~d and 10.6~d, respectively. ZTF18abauprj displays a clear decrease in this ratio with time, with the HV component fading away completely somewhere between $-11.4$ and $-5.8$~d, whereas ZTF19aatlmbo exhibits a far flatter evolution; albeit with large uncertainties on the spectrum at $-10.9$~d. The R$_\text{HVF}$ evolution for these multi-spectra objects is consistent with previous studies showing a decline with phase, with HV components starting out strong and fading away over time \citep{silvermanHVF}. \textcolor{black}{As clear from these two objects, while we see a decrease in R$_\text{HVF}$ with time in individual SNe, this decay occurs at different phases, and we see many multi-spectra targets still exhibiting HVFs after the HV component in ZTF18abauprj had faded away completely. When combining the multi-spectra and single-spectra objects to get a global view of how R$_\text{HVF}$ varies with phase, we find no clear evolution, with some large and small measurements of this ratio and early and late phases, further supporting the idea that the fading away of HV components occurs at different phases in different objects.}

The bottom panel of Fig.~\ref{fig:R_evolution} presents the evolution of the velocity separation, $\Delta v$, with phase. As for R$_\text{HVF}$, the multi-epoch spectra with small temporal differences do not shed much light upon the phase evolution of $\Delta v$. For ZTF18abauprj, we see an upwards evolution and the converse for ZTF19aatlmbo, implying that there is not a singular strict evolutionary trend for such features. However, when looking at the global evolution of the sample, we observe a dearth in the higher velocity separation spectra as we approach maximum light, with all the separations from HVF spectra after $-4$~d clustering below $\sim$5000~\kms. This indicates more separated HV components tend to fade away at earlier phases compared to those that are more entangled with their PV counterpart.

\begin{figure}
 \includegraphics[width = \linewidth]{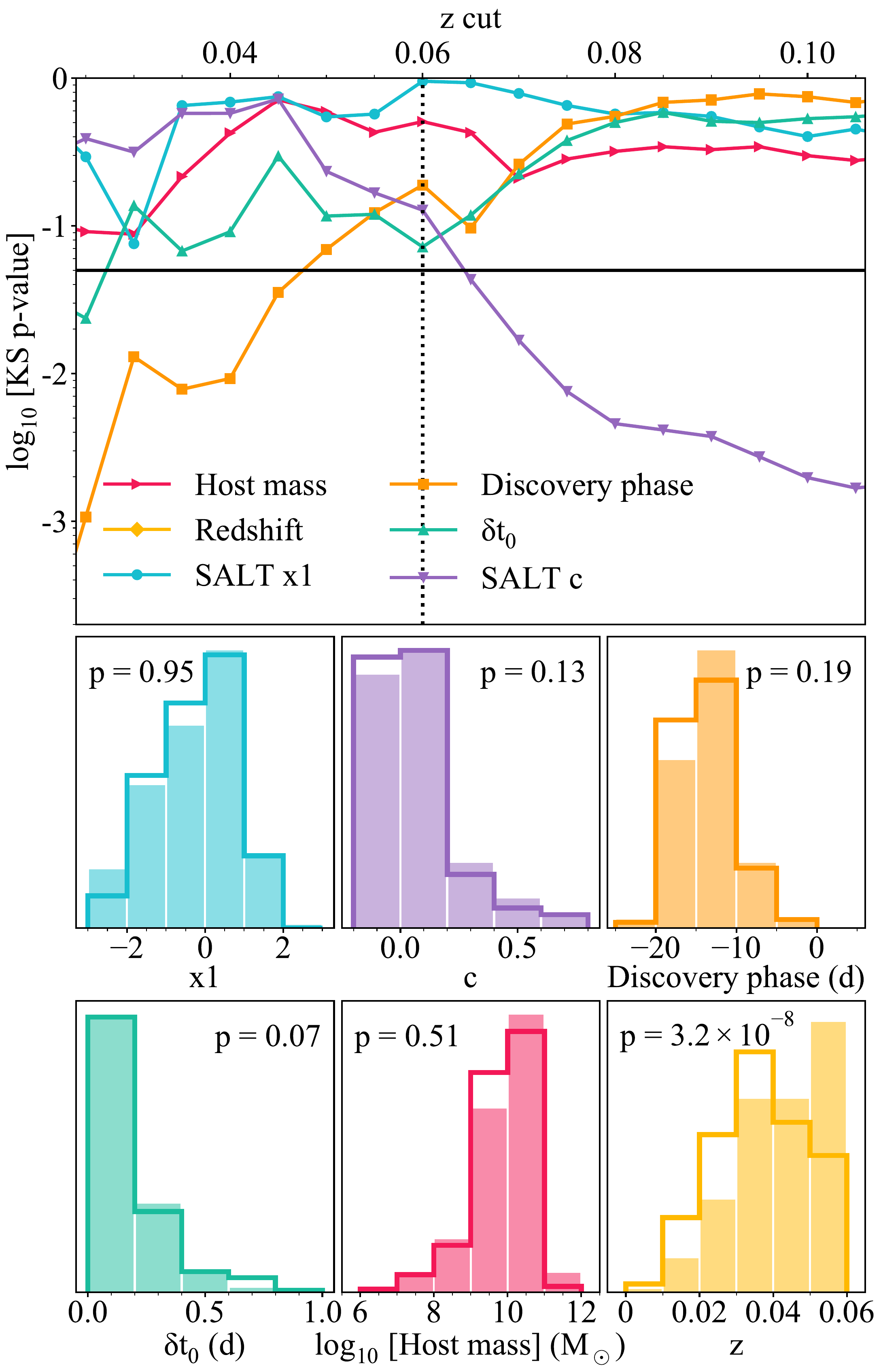}
 \caption{Top: KS p-values between our base sample and the control sample for six parameters. The solid black line describes the threshold below which a parameter should be considered to show a bias when compared to the `low-bias' control sample. The dotted black line indicates the final redshift cut of 0.06. Bottom: Distributions of the control sample (bars) and the final sample of \LowBiasSNe\ SNe after a redshift cut at z~=~0.06 (step). The p-values from the corresponding KS-tests are denoted in each of the panels.}
 \label{fig:results-creating_low_bias_sample}
\end{figure}

\begin{figure}
 \includegraphics[width = \linewidth]{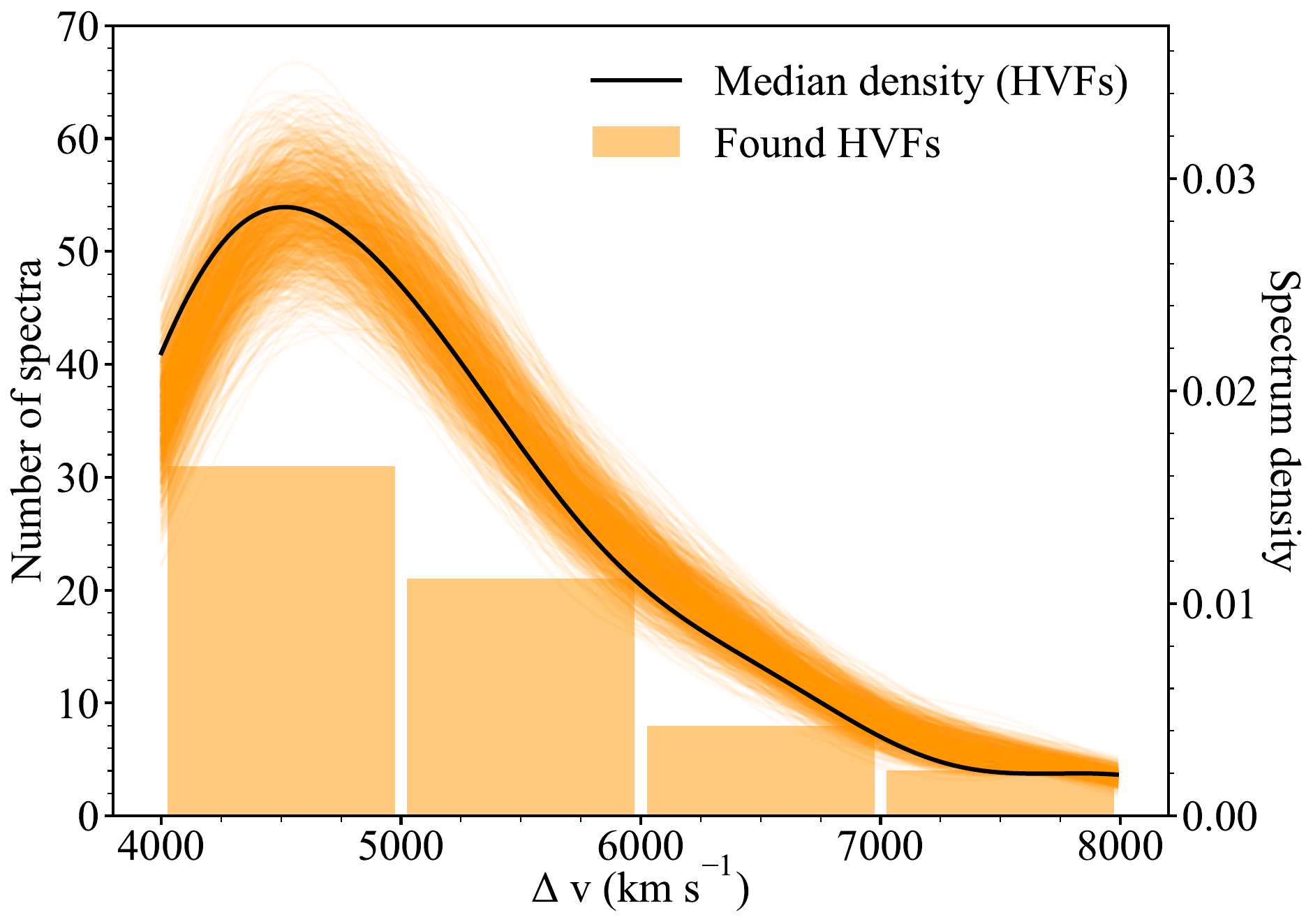}
 \caption{Velocity separation distribution of the \LowBiasHVFSpectra\ HVF spectra from our low-bias sample of \LowBiasSpectra\ SNe. The black curve presents the median spectrum density after correcting for the detection efficiency of our classification method. The individual orange curves correspond to 1000 iterations of sampling the $\Delta v$ values from their individual measurement distributions, rescaling the distribution, and recalculating this density function.}
 \label{fig:vsep_dist}
\end{figure}

\subsection{Defining a low-bias sample}
\label{sec:results:hvf_properties}
In order to analyse global properties of SNe Ia containing \SiFeature\ HVFs, we are required to understand and mitigate the potential biases present in our sample that make it unrepresentative of the global SN Ia population. To this end we define a `control sample' as the 994 SNe Ia from the DR2 volume-limited (\textit{z} $\leq$0.06) sample presented in \cite{mickael_dr2}. At this point we also remove any objects from our sample of \FinalSNe\ SNe that did not pass the initial suggested cuts outlined in Sect.~\ref{sec:data} (fit probability $> 10^{-5}$, $\delta x_1 < 1$, $\delta c < 0.1$, $\delta t_0 < 1$~d), reducing our `base sample' to \BaseBiasedSNe\ SNe Ia. Through Kolmogorov-Smirnov (KS) tests between this base sample and the control sample, we can identify parameters in which our sample exhibits a significant level of bias and then eliminate these by way of a redshift-based cut. KS p-values below the threshold of 0.05 signify that the difference between the two samples is statistically unlikely to have occurred simply by chance, and points to them coming from separation populations.

The top panel of Fig.~\ref{fig:results-creating_low_bias_sample} presents these KS-test results for a number of measurements between the control sample and the base \BaseBiasedSNe\ SNe sample after imposing different redshift cuts, e.g., the data point at $z  = 0.04$, corresponds to the KS-test between the base sample limited to SNe Ia with $z \leq 0.04$ and the full control sample. The parameters investigated here are the SALT2 stretch and colours parameters, $x_1$ and $c$, the phase of first photometric detection, the uncertainty in the date of maximum light ($\delta t_0$), the host galaxy mass (measurements from Smith et al.~in prep), and the host galaxy redshift (not shown on plot). Our base sample of \BaseBiasedSNe\ SNe compared to the control sample has KS p-values that are greater than our bias threshold of 0.05 in all parameters for all redshift cut values, except $c$ and host redshift. As we introduce more restrictive redshift cuts to our sample, we see the KS p-values for $c$ rise, until it crosses our threshold to be considered low bias at a redshift cut of \textit{z}~=~0.06.

The comparison of the redshift distribution between the base and control samples exhibits such small p-values - for all potential redshift cuts - that it falls below the y-axis scaling of the plot in Fig.~\ref{fig:results-creating_low_bias_sample}. This difference is not surprising at redshift cuts either far above or far below that of the control sample (\textit{z}~=~0.06). With a redshift cut equal to that of the control sample we see that the difference between the populations is that our sample peaks at lower redshifts, most likely due to the apparent brightness required to obtain a high enough SNR spectrum pre-peak to enter our sample. The redshift distribution describes the geometric distribution of our objects in space, with the intrinsic physics of the objects described by the other parameters. Therefore, the bias in the redshift distribution  should not impact our results given that we observe no statistically significant bias in the other parameters. We consequently place a cut at \textit{z}~=~0.06, leaving us with \LowBiasSpectra\ spectra from \LowBiasSNe\ SNe from which to investigate global HVF properties.  The distributions of these parameters in this low-bias cut sample can be seen as step plots in the bottom panels of Fig.~\ref{fig:results-creating_low_bias_sample} along with the control sample as the bars.

\subsection{Parameter distributions}
\label{sec:results:parameter_distributions}
In Fig.~\ref{fig:vsep_dist} we present the distribution of velocity separations found in the \LowBiasHVFSpectra\ HVF spectra of the low-bias sample as the solid bar histogram. In general we observe a distribution skewed towards smaller velocity separations with very few spectra showing the extreme separations of 8000~\kms. As seen in Fig.~\ref{fig:true_false_positive_rates} our classification method is very sensitive to the larger velocity separations, however its detection efficiency drops off as the features encroach on one another and become increasingly entangled. As an example, we have a true-positive rate of $\sim$25\%\ for features with SNR 8 and \textcolor{black}{dispersion} 2~\AA/pix that possess a velocity separation of 4000~\kms, implying that for every one spectrum that we identify with this parameter set, we miss three others; the true value is the number of identified targets divided by the fractional true-positive rate. Therefore, we can draw efficiency values from our 3D GP interpolation of the true-positive rates to scale the distribution of velocity separations on a spectrum-by-spectrum basis. This is first performed using the measured values of velocity separation, for which a Gaussian KDE of the probability density function (PDF) can be seen in Fig.~\ref{fig:vsep_dist} as the black solid line. In order to probe the potential variation of this distribution we employ a Monte-Carlo method, sampling each of the velocity separations in the sample from their individual distributions - described by their measured medians and upper and lower uncertainties - before then performing the detection efficiency scaling with the resulting $\Delta v$ values. This procedure is repeated 1000 times with each of these PDFs plotted as a faint orange line in Fig.~\ref{fig:vsep_dist}. These PDFs are all multiplied by the number of spectra that they represent and therefore, represent spectrum densities instead of probabilities.

\begin{figure}[b]
 \includegraphics[width = \linewidth]{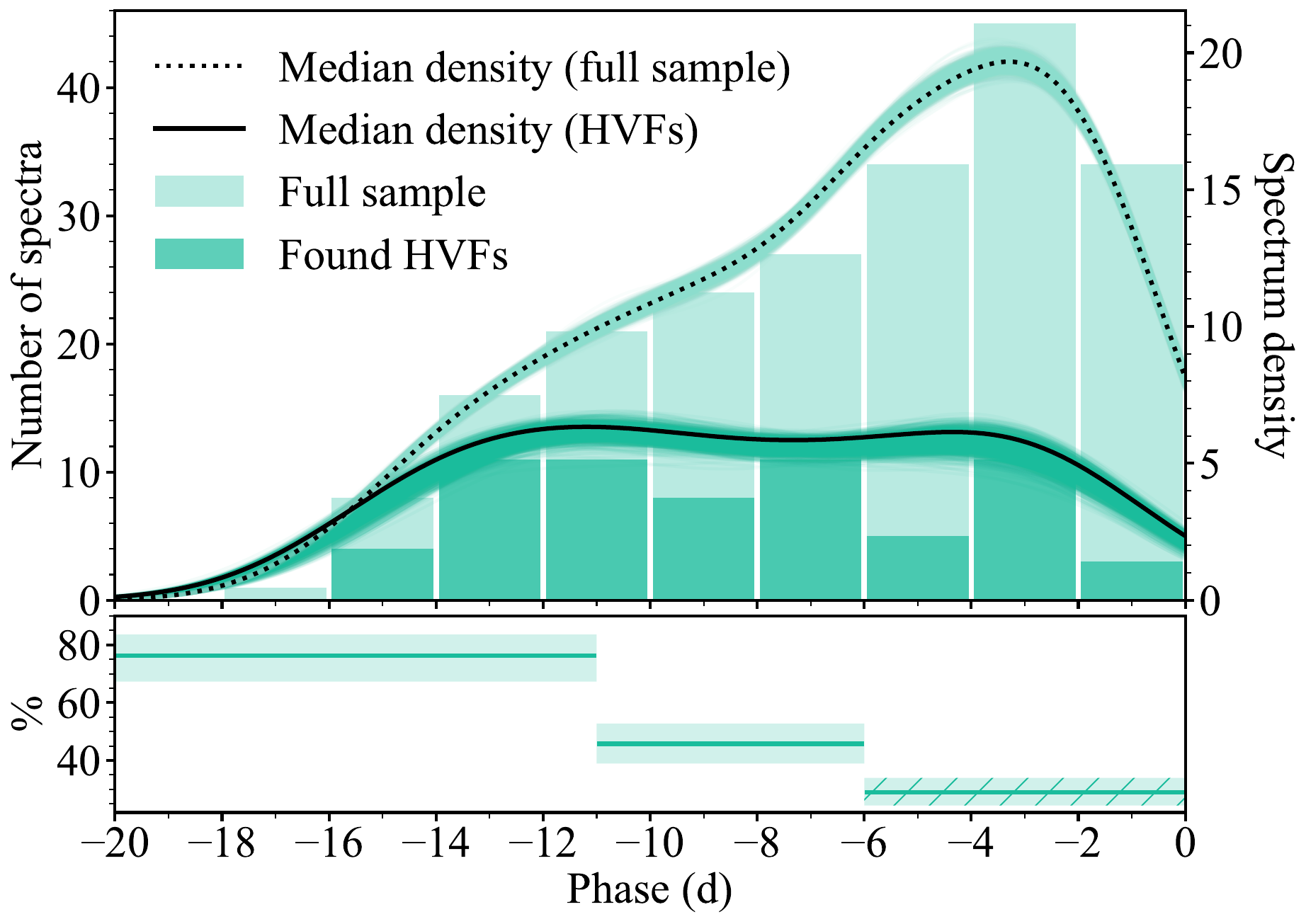}
 \caption{Phase histogram for the \LowBiasHVFSpectra\ HVF spectra compared to the full \LowBiasSpectra\ low-bias spectral sample. The solid black curve presents the median HVF spectrum density after correcting for the detection efficiency of our classification method, with the dotted black curve as the density of the full \LowBiasSpectra\ spectra. The individual green curves (dark for HVF and pale for the full sample) correspond to 1000 iterations of sampling the $\Delta v$ and phase values from their individual measurement distributions, rescaling the HVF distribution, and recalculating these density functions.}
 \label{fig:phase_dist}
\end{figure}

In order to probe the potential variation of this distribution we employ a Monte-Carlo method, sampling each of the velocity separations in the sample from their individual distributions - described by their measured medians and upper and lower uncertainties - before then performing the detection efficiency scaling with the resulting $\Delta v$ values. This procedure is repeated 1000 times with each of these PDFs plotted as a faint orange line in Fig.~\ref{fig:vsep_dist}. \textcolor{black}{In each iteration we account for the variation due to false positives by randomly reclassifying some HVF spectra with $\Delta v<5000$~\kms\ as non-HVF in accordance with our adopted conservative false positive rate of 2\%.} These final PDFs are all multiplied by the number of spectra that they represent and therefore, represent spectrum densities instead of probabilities.

\begin{figure}[b]
 \includegraphics[width = \linewidth]{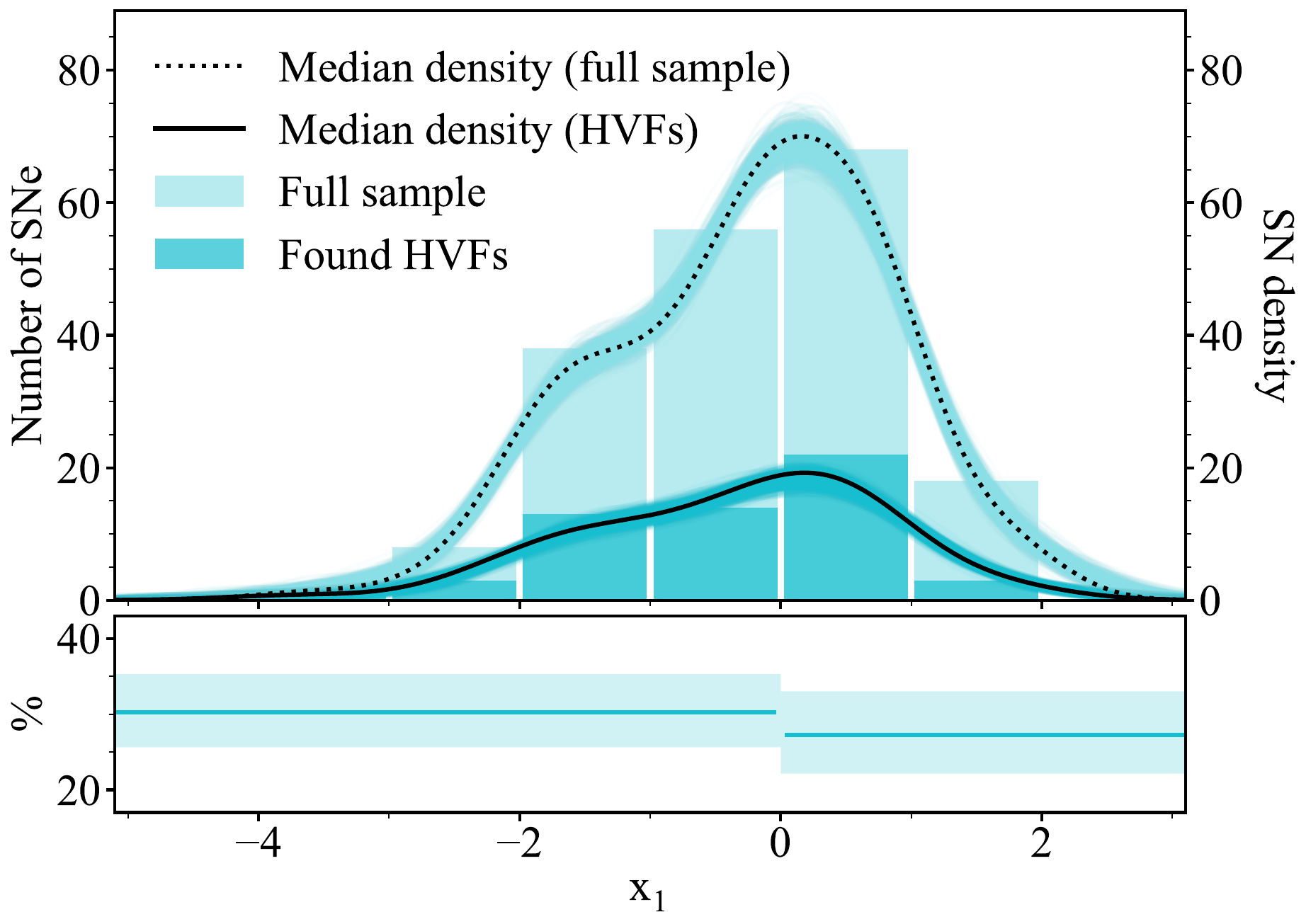}
 \caption{SALT2 x$_1$ histogram for the \LowBiasHVFSNe\ HVF SNe compared to the full \LowBiasSNe\ low-bias sample. The solid black curve presents the median HVF SN density, with the dotted black curve as the density of the full \LowBiasSpectra\ spectra. We perform no detection efficiency scaling upon the x$_1$ parameter distribution. The individual blue curves (dark for HVF and pale for the full sample) correspond to 1000 iterations of sampling the x$_1$ values from their individual measurement distributions and recalculating these density functions.}
 \label{fig:x1_dist}
\end{figure}

Figure~\ref{fig:phase_dist} presents the phase distribution of the spectra displaying HVF against our full low-bias sample. The full sample is presented as the faint bars with the HVF spectra as the solid bars. As before, we perform the detection efficiency scaling on a spectrum-by-spectrum basis and plot the resulting density function from the velocity separation values as the solid black curve. The spectrum density function of the full phase distribution is plotted as the dotted black line. As for the velocity separation analysis, we perform Monte-Carlo sampling for these two phase distributions, \textcolor{black}{randomly reclassifying some HVF spectra with $\Delta v<5000$~\kms\ as non-HVF in accordance with the false positive rate and} drawing velocity separations and phases from the corresponding individual distributions to calculate new PDFs. In each of these 1000 iterations we also integrate over three phase bins to calculate the percentage of spectra exhibiting a \SiFeature\ HV component. These three regions are divided up to each hold $\sim$1/3 of the \LowBiasHVFSpectra\ HVF spectra from the low-bias sample, giving bins for $<-$11 d, $-11$ to $-6$ d and $>-$6 d. The percentages of HVF spectra for these three phase ranges are shown in the bottom panel of Fig.~\ref{fig:phase_dist} along with the shaded regions representing 68\%\ confidence intervals calculated via the Clopper-Pearson method. These calculations indicate that we find \SiFeature\ HV components in 76$\pm^{7}_{9}$\%\ of spectra before $-11$~d, 46$\pm{7}$\%\ of spectra between $-11$ and $-6$~d, and 29$\pm{5}$\%\ of spectra between $-6$~d and maximum light. If we were to use the original detection efficiencies before introducing the cuts  to only consider simulations with $a_\text{HV}\leq0.25$ and $c_\text{HV}\leq70$~\AA\ (see Section~\ref{sec:results:validation_priors}), we calculate slightly higher percentages of 83$\pm^{6}_{8}$\%, 49$\pm{7}$\%, and 32$\pm{5}$\%\ for these three phase intervals. Therefore, \SiFeature\ HV components are close to $\sim$2.5 times more common at early phases, but still appear in approximately one third of spectra in the week before maximum light.

\subsection{Light curve observables and host measurements}
\label{sec:results:observables}

In Fig.~\ref{fig:x1_dist} we present the distributions of the SALT2 light curve parameter $x_1$ for the SNe Ia in our full low-bias sample and only those showing evidence of HVFs as the faint and solid histograms, respectively. This information is then displayed in the form of SN density functions for the full sample (dotted) and those for which we have a spectrum with an identified HVF (solid). As for the $\Delta v$ and phase parameters before, we perform a Monte-Carlo sampling of the individual $x_1$ measurements over 1000 iterations to assess the amount of variation in these density curves \textcolor{black}{accounting for the estimated false positive rate and uncertainties in the $x_1$ values}. For each of these iterations, we integrate under the two density curves over two regions ($x_1<0$ and $0\leq$ $x_1$) and calculate the percentage of SNe exhibiting HV components in the \SiFeature. As visible in the bottom panel of Fig.~\ref{fig:x1_dist}, these two percentages (30$\pm{5}$\%\ and 27$\pm^{6}_{5}$\%) are consistent with one another and support the idea of HVFs being ubiquitous across the SN Ia population. As before for the phase percentages, these uncertainties represent 68\%\ confidence intervals.

\textcolor{black}{As seen in Fig.~\ref{fig:phase_dist}, \SiFeature\ HVFs are more prevalent at earlier phases and there is a significant decrease in the percentage of SN Ia spectra displaying HVF features as the phase approaches maximum light. By cutting the sample and repeating these percentage calculations with only the earliest phases, we can test the impact of this phase dependence on any potential trend with $x_1$. We introduce a phase cut at $-8.7$~d - the median HVF spectral phase - and recalculate the rates of occurrence in HVF in low and high $x_1$ SNe Ia at these earlier phases as 43$\pm^{11}_{10}$\%\ ($x_1<0$) and 56$\pm{11}$\% ($0\leq$ $x_1$). While the difference between these two percentages is slightly larger than before (and in the opposite direction), they still overlap in the 1$\sigma$ uncertainty region and therefore support the idea of HVF ubiquity.}

\begin{figure}
 \includegraphics[width = \linewidth]{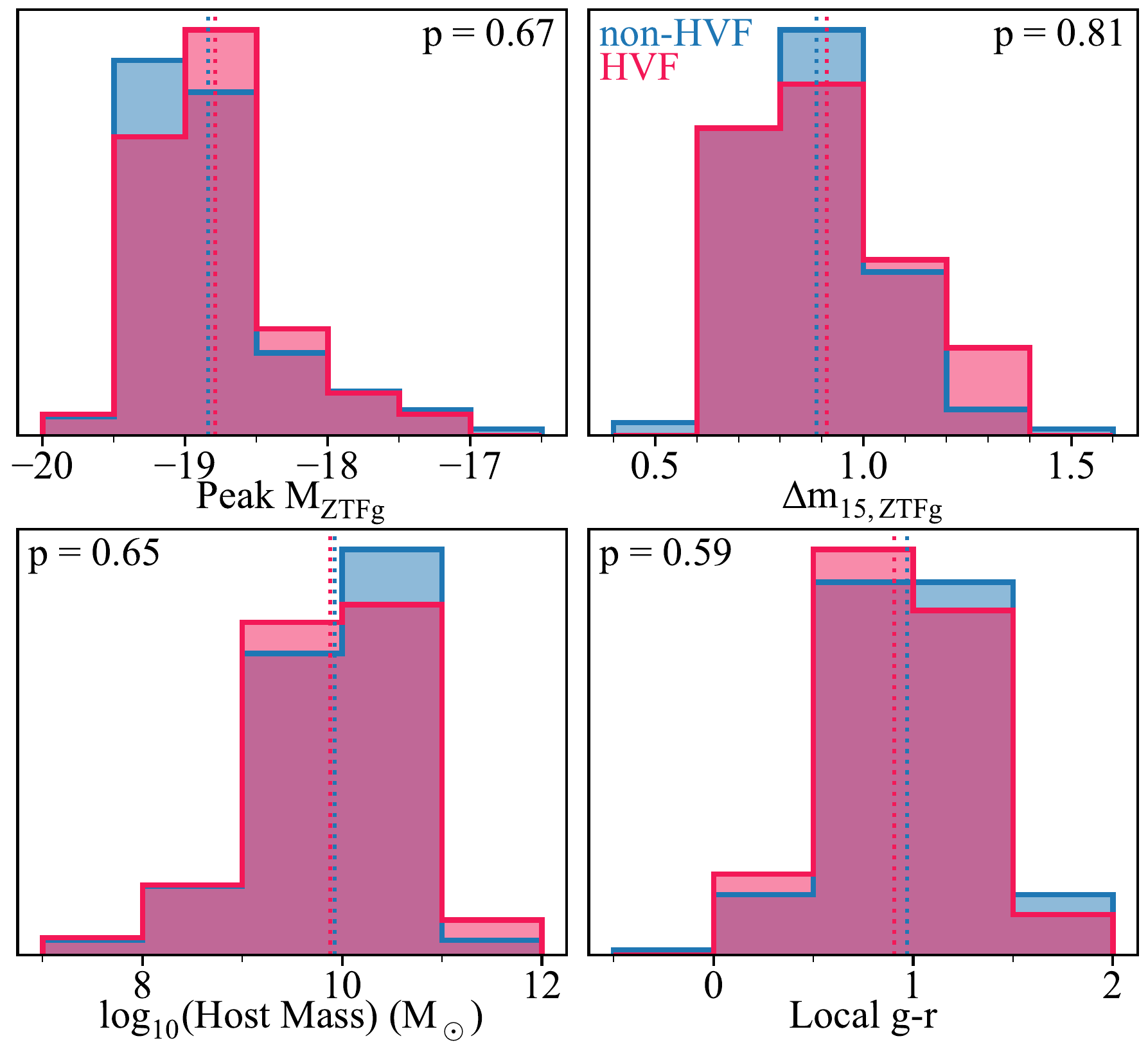}
 \caption{Distributions of peak ZTF\textit{g}-band magnitude, $\Delta m_{15}$ in ZTF\textit{g}, host galaxy mass, and local g$-$r colour for the HVF and non-HVF subsets of the \LowBiasSNe\ SNe low-bias sample. The means of each distribution are displayed as the dotted lines with the KS test p-values indicated in each panel.}
 \label{fig:observables}
\end{figure}

We also investigated the relationship between the presence of HVFs and light-curve parameters (peak absolute ZTF\textit{g}-band magnitude, the \textit{g}-band decline rate in 15 days post maximum, $\Delta$m$_{15, ZTFg}$) measured from GP fitting in \cite{georgios_dr2}. These are shown in the top panels of Fig.~\ref{fig:observables}. We employ KS tests to quantify the likelihood of the two populations hailing from the same parent population. Resulting p-values that fall below the threshold of 0.05 indicate that the difference between the two samples for a given measurement is unlikely to have occurred simply by chance, and may point to a physical underlying distinction. We observe no significant difference between the HVF and non-HVF populations in terms of the peak ZTF\textit{g}-band absolute magnitude and $\Delta$m$_{15}$ parameters with the p values as 0.67 and 0.81 respectively. This again supports the ubiquity of these features across the Ia population.

The host galaxy mass and local \textit{g-r} colour (Smith et al.~in prep.) are shown in the bottom panels of Fig.~\ref{fig:observables}. The p values for the host mass and local colour distributions of the HVF and non-HVF samples are large relative to the 0.05 threshold, with values of 0.65 and 0.59, respectively, again consistent with a lack of a difference in the local environment between SNe Ia with and without a \SiFeature\ HVF.

\textcolor{black}{As before for investigating trends with $x_1$, we repeated these statistical tests exclusively with spectra before $-8.7$~d. We once again find large p-values for peak absolute magnitude and $\Delta$m$_{15}$ in the ZTF\textit{g}-band (0.91 and 0.52) as well as for the host galaxy mass and local \textit{g-r} colour (1.0 and 0.94), indicating no difference between the HVF and non-HVF populations in these observables.}


\section{Discussion}
\label{sec:discussion}

In Section~\ref{sec:discussion:silverman}, we discuss our results in the context of the results of the literature, particularly those of \cite{silvermanHVF}. In Section~\ref{sec:discussion:wang_reclass} we investigate the potential false classification rates of the `Wang' \citep{wang_classification} and the `Branch' \citep{branch2006,branch2009}, classification schemes if HVFs are not taken into account. Finally, we compare our velocity distributions to hydrodynamical explosion models in Section~\ref{sec:discussion:hvf_origins}.

\subsection{Comparison to literature}
\label{sec:discussion:silverman}
\cite{silvermanHVF} investigated the properties of SNe Ia both with and without HVF in the context of the velocity-based classification scheme of \cite{wang_classification}. The `Wang classification scheme' divides objects into normal velocity (NV\wang) and high velocity (HV\wang) subclasses depending on whether the photospheric velocity (measured from the \SiFeature\ feature) is less than or greater than 11800~\kms\ around peak ($-$5 to 5~d), respectively. We chose to add the `W' subscript here to differentiate between the high-velocity Wang subclass (HV\wang) and the high-velocity components (HV). While the cutoff of 11800~\kms\ was employed by \cite{silvermanHVF}, we chose to adopt a cutoff of 12000~\kms\ for consistency with other ZTF DR2 studies \citep{burgaz_dr2_spectra}. Our Wang classifications are drawn from the latest spectrum we have for each object. In the case of the HV\wang\ class, only spectra later that $-$5~d were used. However, for any object for which the latest pre-peak spectrum has PV~$<12000$~\kms, regardless of phase, we denote it as NV\wang\ as the photospheric velocity is not expected to increase and the objects will still have PV$<12000$~\kms\ around peak. These classifications are then augmented by classifications from \citep{burgaz_dr2_spectra} for which they examined a slightly later phase range of $-$5 to 5~d with respect to peak brightness. 

\begin{figure}
 \includegraphics[width = \linewidth]{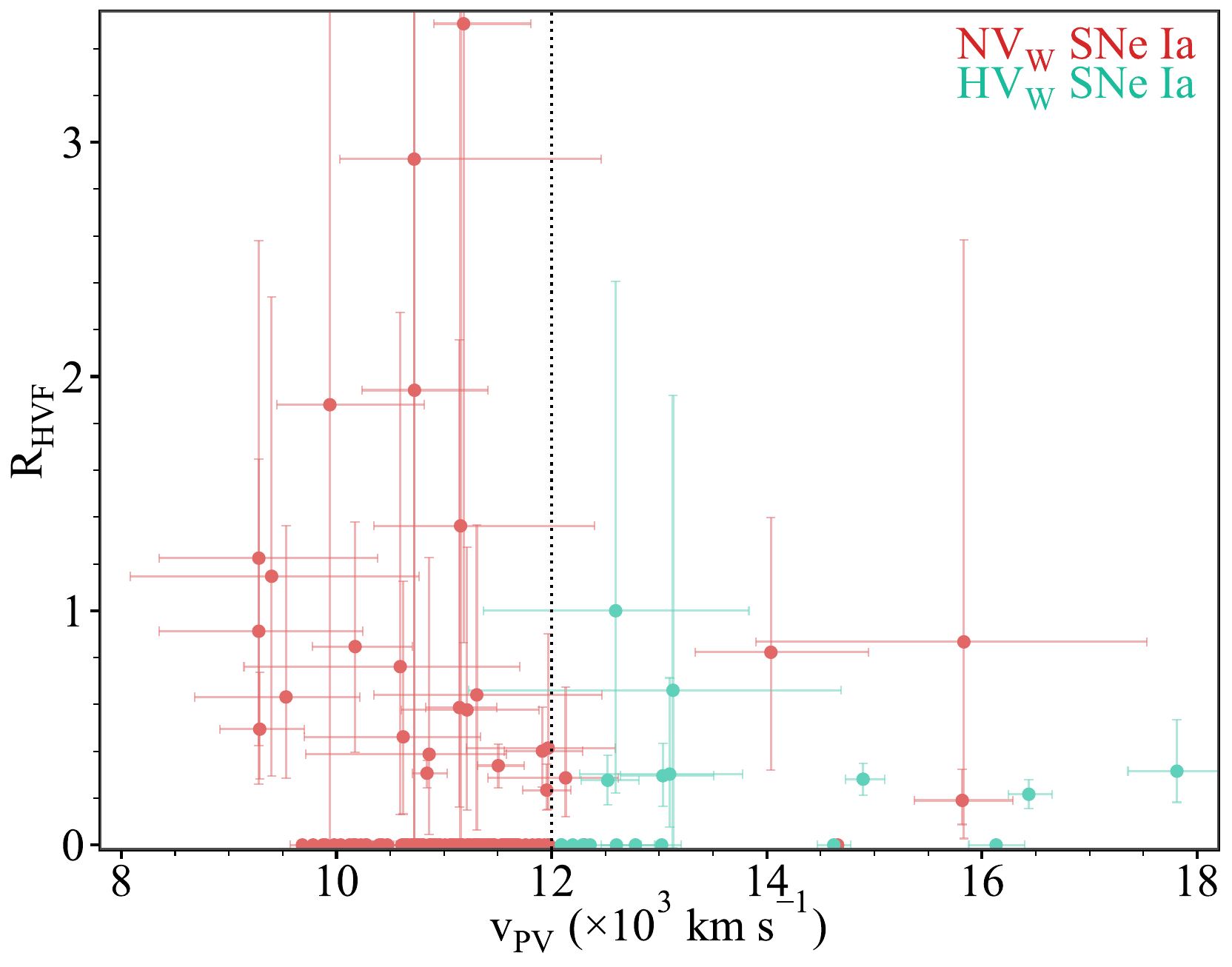}
 \caption{The ratio of the pseudo-equivalent widths of the HV and PV components against the photospheric velocity of the \SiFeature\ feature for the low-bias sample of \LowBiasSNe\ SNe. For objects with multiple spectra we take the mean value in both dimensions, treating the uncertainties with the \textsc{asymmetric uncertainty} package \protect{\citep{asymmetric_uncertainties}}. The vertical dotted line indicates the 12000~\kms\ cut velocity at maximum light between the NV\wang\ (red) and HV\wang\ (green) SNe Ia.}
 \label{fig:R_against_PV}
\end{figure}

Figure~\ref{fig:R_against_PV} presents the pEW ratio, R$_\text{HVF}$, against the photospheric velocity of the \SiFeature\ feature for the low-bias sample of \LowBiasSNe\ SNe Ia. When investigating the same parameters, \cite{silvermanHVF} found a dearth of NV\wang\ SNe (PV$<12000$~\kms) exhibiting HVFs in the \SiFeature\ (R$_\text{HVF}$ $>$0), with these features found more predominantly in HV\wang\ SNe. Contrary to this, we find HVFs in 26 of the NV\wang\ classified objects in our low-bias sample (36 in the full sample), largely populating the empty region of the parameter space seen by \cite{silvermanHVF}. An increase in R$_\text{HVF}$ with increasing photospheric velocity was also observed by \cite{silvermanHVF}. While this correlation makes intuitive sense for the evolution of individual objects, with R$_\text{HVF}$ and photospheric velocity both decreasing over time, we observe no such global correlation in Fig.~\ref{fig:R_against_PV}. As before with the lack of a global downwards trend in the \textcolor{black}{middle} panel of Fig.~\ref{fig:R_evolution}, this indicates that the onset of the R$_\text{HVF}$ evolution begins at different phases for different objects.

\subsection{Wang and Branch reclassifications}
\label{sec:discussion:wang_reclass}

As described above, Wang classifications (NV\wang\ vs. HV\wang) are generally measured using the \SiFeature\ photospheric velocity in the phase range $-$5 to 5~d with respect to peak. However, the specific measurement of the photospheric velocity will change depending upon whether we employ a single- or a double-velocity component model. Historically this feature has been treated as a single component around peak, believed to be free of any HV components. However, our study, as well as the results from \cite{silvermanHVF}, suggests that HV components might be more common at these phases (later than $-$5 d) than previously thought, with one third of SN Ia spectra between $-$5 d and peak displaying a HVF. Therefore, we pose the question, if in these large spectroscopic studies we were to consider and account for HV \SiFeature\ absorption close to maximum light, what percentage of the HV\wang\ classifications would be overturned in favour of a NV\wang\ subtype?

\begin{figure}
 \includegraphics[width = \linewidth]{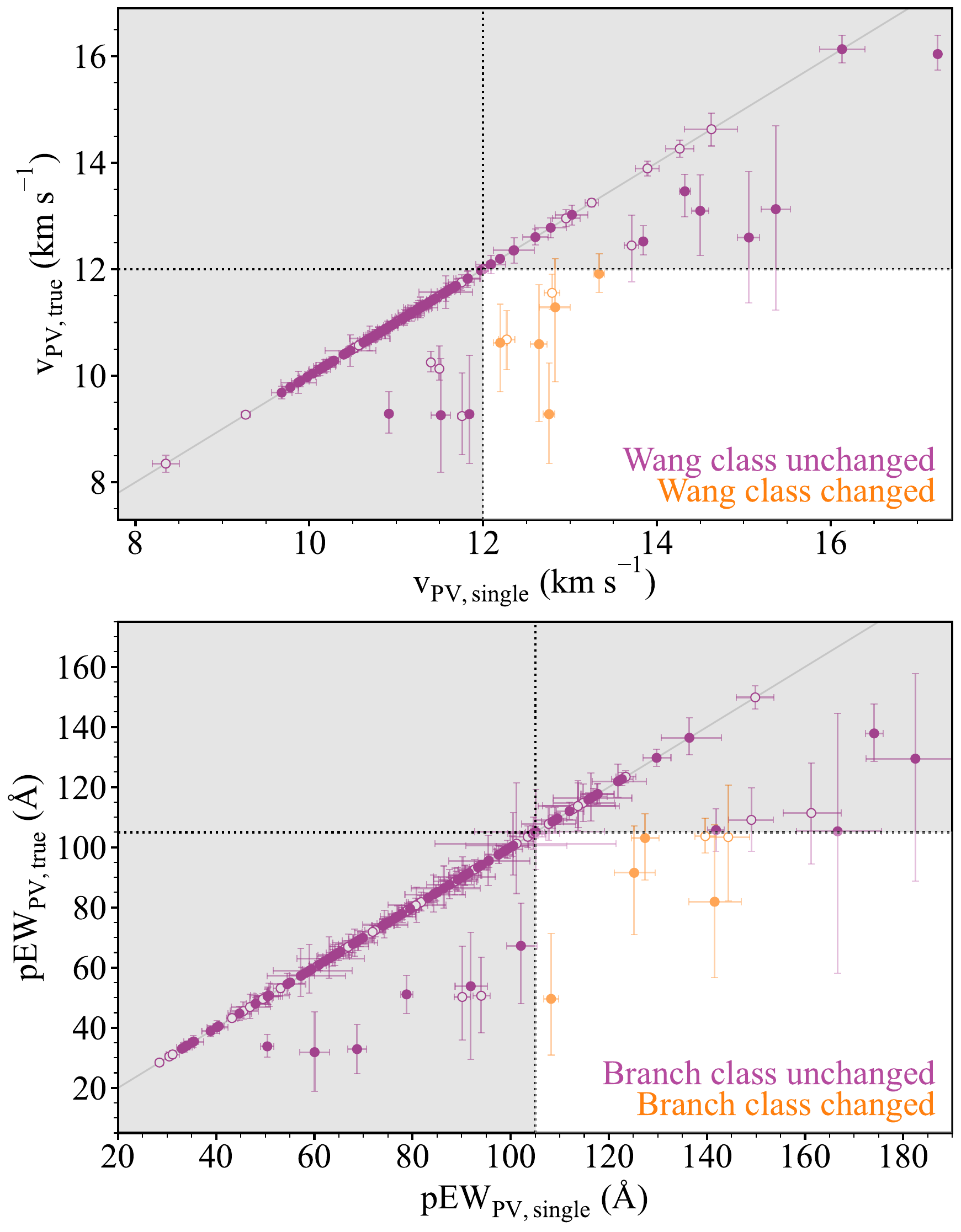}
 \caption{Comparison of the photospheric velocity (top) and pEW (bottom) measured using the single component fits against taking into consideration the HV components and using the two component fits wherever a HV component was identified. \textcolor{black}{Solid datapoints represent the low-bias sample, with hollow datapoints corresponding to the remaining objects from the full sample with z~>~0.06.} The points indicated in orange would be misclassified in the Wang scheme and Branch scheme in the top and bottom panels respectively, if the HV components were not to be considered. These two classification schemes concern spectra around maximum light, as such all spectra presented here have phases greater that $-$5~d.}
 \label{fig:changed_bl-wang_classifications}
\end{figure}

In the top panel of Fig.~\ref{fig:changed_bl-wang_classifications}, we compare the photospheric velocity from the single component fits against the `true' photospheric velocities (i.e. single component fits for spectra without a HVF, and double component fits for spectra with a HVF). All spectra identified to have a HV component exhibit lower photospheric velocities in their two-component fits as would be expected. We indicate the Wang classification cut off velocity of 12000~\kms\ by the vertical and horizontal dotted lines, highlighting the region in the bottom right in which the objects would be classified as HV\wang\ with the single component model, but as NV\wang\ by the double-component model (orange points). When considering the 85 objects from the low-bias sample which possess a spectrum later than $-$5~d, there are five SNe Ia (seven of the 150 SNe from the full sample with a spectrum later than $-$5~d), indicating that $26\pm^{14}_{11}$\%\ of HV\wang\ classifications ($24\pm^{11}_{8}$\%\ for the full sample) before peak would be incorrect if we were to not consider HV components. These uncertainties represent 68\%\ confidence intervals and are calculated as binomial with the Clopper-Pearson interval. Such large uncertainties are the result of low number statistics and while this leaves the percentages fairly unconstrained, this suggests that HVF components could cause a significant rate of false HV\wang\ classifications in the 5~d before maximum light. 

We similarly perform this analysis with respect to the Branch classification scheme \citep{branch2006, branch2009}, which divides up the SN Ia population based upon the equivalent widths of the \SiFeature\ and \SiFeatureOther\ lines. Broad line SNe Ia (BL) exhibit pEW$_{\text{Si 6355}}>105$ \r A with a pEW$_{\text{Si 5972}}<30$ \r A. While we have no information on the \SiFeatureOther, we can examine the effects of HV components upon the measured \SiFeature\ pEW. The bottom panel of Fig.~\ref{fig:changed_bl-wang_classifications} displays the measured photospheric pEW coming from the single component fits against the `true' photospheric pEWs (as above for the Wang analysis). The cutoff width at 105 \r A is indicated by the dotted lines, with the orange points once again corresponding to those spectra that would receive an incorrect classification if we were to not consider the HV components. In the Branch scheme parameter space there are four SNe that are misclassified (six for the full sample), resulting in $20\pm^{13}_{9}$\%\ of incorrect BL classifications ($18\pm^{9}_{7}$\%\ for the full sample) with respect to the \SiFeature\ line. In their sample of BL SNe Ia, \cite{2019ein_Yarbrough} also found the BL population to have higher \SiFeature\ velocities than their CN counterparts, further supporting this idea as the HV components would also cause an offset in measured single component velocities to higher values. These uncertainties represent 68\%\ intervals as before for the Wang reclassifications, and once again the low number of datapoints results in loosely constrained percentages.

\textcolor{black}{With generally smaller velocity separations in this phase range around peak, false positives are more probable and could have the inverse effect in both these cases. Incorrectly employing a double-component model to a \SiFeature\ feature with no HVF would artificially decrease the photospheric velocity and width, potentially resulting in HV\wang\ SNe being classified as NV\wang\ and Branch BL SNe being missed. Therefore, the misclassification rates calculated here represent upper limits.}

\subsection{HVF origins}
\label{sec:discussion:hvf_origins}

\begin{figure}
 \includegraphics[width = \linewidth]{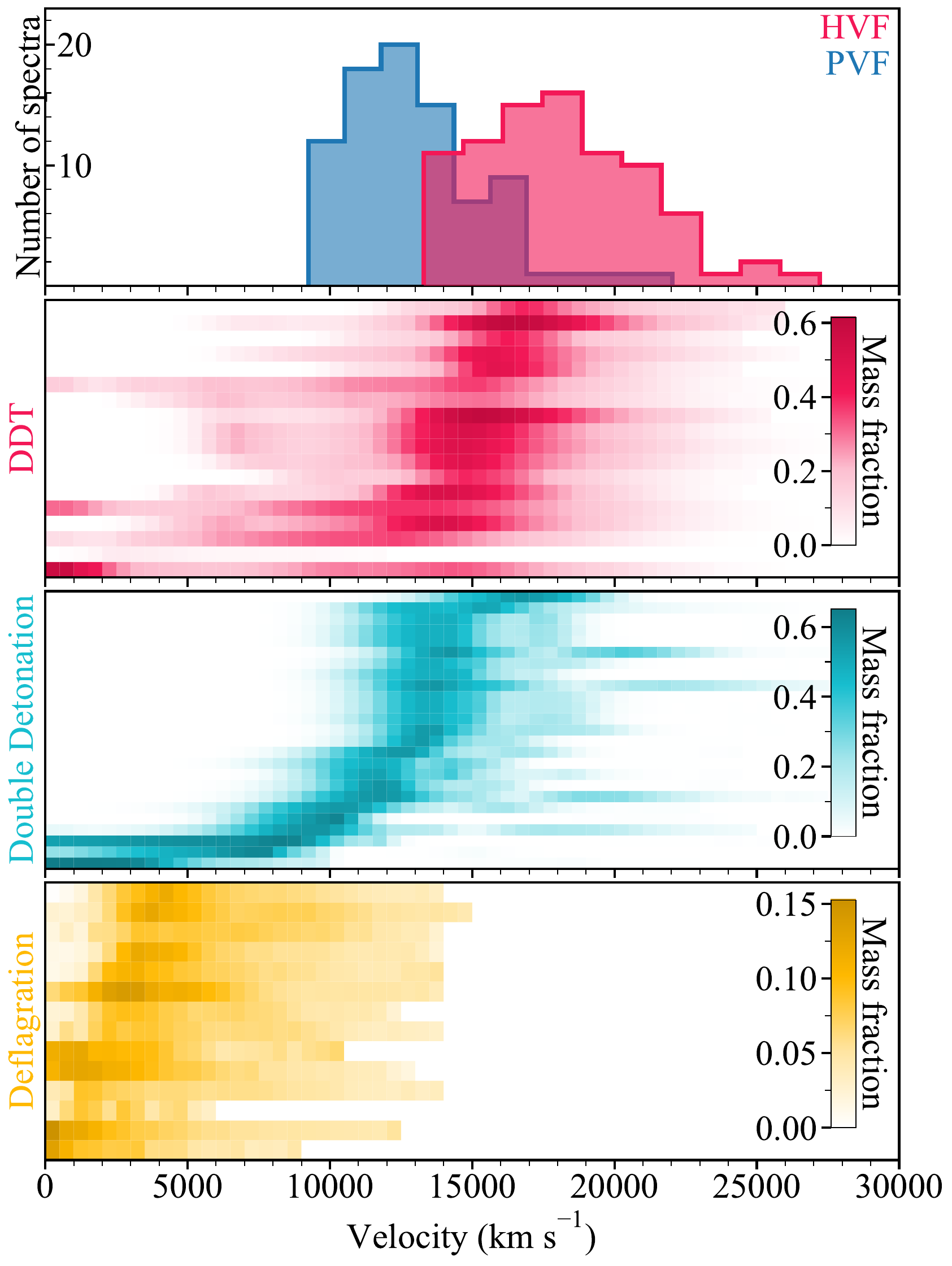}
 \caption{Comparison of the measured distributions of PV and HV components in our sample for spectra identified as having both (top) against the distributions of silicon from a number of theoretical explosion models from the HESMA archive \citep{hesma}. Each of the three panels corresponds to a different explosion mechanism, with each row of the colour plot as an individual, unique model.}
 \label{fig:explosion_models}
\end{figure}

While the specific origins and formation channels of these HV components of the \SiFeature\ feature remain unclear, they provide clear evidence of intermediate-mass element material in the upper ejecta of the majority of SNe Ia. Regardless of the explosion mechanism or progenitor channel, any model aiming to reproduce these features will require some fraction of silicon at the corresponding high velocities.

In Fig.~\ref{fig:explosion_models} we present the measured velocity distributions for the PV and HV components for the HVF classified spectra in our sample, with the Si-abundance profiles from a number of explosion models from the Heidelberg Supernova Model Archive \citep[HESMA;][]{hesma}, arising from three different explosion mechanisms. In pink we plot the delayed-detonation (DDT) models (\citealt{ropke_ddt}; \citealt{ivo_ddt}; \citealt{2014_ddt}) arising from an initial subsonic deflagration that later transitions to a supersonic detonation. In blue, we plot the double-detonation models (\citealt{2012_doubledet}; \citealt{2020_doubledet}; \citealt{2021_doubledet}), which involves a detonation in a helium shell atop the white dwarf, sending a shock-wave inwards which converges in - and subsequently detonates - the core. Finally, in yellow we show the silicon abundances from the pure subsonic deflagration models (\citealt{2014_def}; \citealt{2015_def}).

As expected, the pure deflagration models are very homogeneous in nature, possessing very flat distributions of the elements in the model. As a result of the low energy they also do not produce material moving as velocities faster than around 15000~\kms, and therefore, are incapable of producing the HV - and many of the PV - components that we see in the DR2. The DDT and double-detonation models possess more layered ejecta structures, with peaks in the silicon abundance profiles in the 10000 to 15000~\kms\ range. The majority of these models also exhibit silicon at velocities covering the full HVF velocity distribution. The formation of such HVFs is dependant upon the ionisation profile, which in turn is dependant upon not only the abundances, but also the density profile of the plasma. This will be the focus of a follow-up modelling study with a handful of well-sampled HVF objects (Harvey et al. in prep.).

We also note here that the models presented in Fig.~\ref{fig:explosion_models} are designed to reproduce the overall SN Ia photometric and spectroscopic evolution coming from the primary white dwarf itself. These HVFs may instead be the result of environmental components such as CSM, which are not included in the models. Through the spectroscopic modelling of the HVFs exhibited by SN~1999ee, \cite{mazzali_99ee} explored the ejecta conditions needed to recreate the evolution. In terms of an abundance enhancement, they required a region of the upper ejecta to be dominated by silicon and calcium ($\sim$90 and $\sim$10 per cent respectively), which is hard to justify in terms of nuclear burning. Therefore, they concluded that the density enhancements in these regions to be a more plausible reason for the HVFs. In \cite{mazzali2005} they suggested the sweeping up of surrounding CSM to be responsible for the line-forming region higher up in the ejecta, pointing to a thick disk and/or high density CSM environment produced by a companion wind. They also proposed angular fluctuations as a potential cause of density enhancement, with HVFs being the result of three-dimensional effects. This was further supported by \cite{Tanaka_3d_hvf} who investigated various density enhancement geometries and degrees of photospheric coverage to produce the observed HVF population. This multidimensional requirement is supported by the observed polarisation difference between the HV and PV components of the Ca NIR triplet in SN~2001el (\citealt{wang2003}; \citealt{kasen2003}) and in the \SiFeature\ of SN~2019ein/ZTF19aatlmbo \citep{2019ein_Patra}


\section{Conclusions}
In this work we conducted a thorough search for \SiFeature\ high-velocity components within the ZTF Cosmology Data Release 2 by way of MCMC fitting and model selection with the Bayesian Information Criterion. Through simulations, we quantified the detection efficiency over the spectral quality (SNR and \textcolor{black}{dispersion}), as well as for simulated velocity separation between the two components. These simulations were also employed to test the accuracy of the estimated uncertainties and identify and correct the biases in the model parameters. Within our sample of \FinalSpectra\ spectra we identified \HVFSpectra\ \SiFeature\ HV components. We subsequently defined a low-bias subsample with a redshift cut at 0.06 (\LowBiasSpectra\ spectra from \LowBiasSNe\ SNe) which was then used to calculate rates. Conclusions from the analysis are as follows:
\begin{enumerate}
    \item 76$\pm^{7}_{9}$\%\ of spectra before $-$11~d exhibit HVF in the \SiFeature\ feature. This percentage drops as we approach maximum light with 29$\pm{5}$\%\ of spectra showing HVFs after $-6$~d.
    \item For individual objects, we observe a decrease in the HVF strength ratio R$_\text{HVF}$ with time, corresponding to the fading away of the HV component to leave just the PV absorption. This evolution occurs at different phases for different objects and therefore, globally we see a relatively uniform distribution of R$_\text{HVF}$ with phase.
    \item HV components with larger separations from their PV counterparts fade at earlier epochs, leaving just the smaller separations as we approach maximum light. We do not observe consistent $\Delta v$ evolution from object to object, with some increasing, others decreasing, and some remaining constant.
    \item We see no tendency of \SiFeature\ HVFs to occur in objects with high or low SALT $x_1$ values, supporting the proposed ubiquity of HVFs across the SN Ia class.
    \item Comparison of a low-bias subsample of our targets with the volume-limited DR2 sample (defined as a control sample) shows no statistical difference between the HVF and non-HVF populations in ZTF\textit{g}-band peak magnitude and $\Delta m_{15}$(ZTF\textit{g}), host mass, or local host \textit{g-r} colour.
    \item We do not reproduce the finding from \cite{silvermanHVF} that Wang classification scheme HV\wang\ objects are more likely to exhibit \SiFeature\ HVFs, with a fairly significant portion of our HVF spectra coming from NV\wang\ objects.
    \item  The fitting of \SiFeature\ profiles without consideration of contributions from potential HV components may cause a number of false HV\wang\ classifications. We calculate \textcolor{black}{an upper limit for} the false HV\wang\ classification rate of $26\pm^{25}_{17}$\%\ in the phase range $-$5 to 0~d.
    \item Similarly for the Branch classification scheme, we find that \textcolor{black}{up to} $20\pm^{24}_{14}$\%\ of broad line (BL) classifications drawn from spectra in the phase range $-$5 to 0~d are the result of HV components making for broader line profiles.
\end{enumerate}

\begin{acknowledgements}
The research conducted in this publication was funded by the Irish Research Council under grant number GOIPG/2020/1387. K.M., U.B., G.D., M.D., and J.T. acknowledge support from EU H2020 ERC grant no. 758638. L.G. acknowledges financial support from the Spanish Ministerio de Ciencia e Innovaci\'on (MCIN), the Agencia Estatal de Investigaci\'on (AEI) 10.13039/501100011033, and the European Social Fund (ESF) "Investing in your future" under the 2019 Ram\'on y Cajal program RYC2019-027683-I and the PID2020-115253GA-I00 HOSTFLOWS project, from Centro Superior de Investigaciones Cient\'ificas (CSIC) under the PIE project 20215AT016, and the program Unidad de Excelencia Mar\'ia de Maeztu CEX2020-001058-M. Y.-L.K. has received funding from the Science and Technology Facilities Council [grant number ST/V000713/1]. This project has received funding from the European Research Council (ERC) under the European Union's Horizon 2020 research and innovation programme (grant agreement n°759194 - USNAC).
This work has been supported by the research project grant “Understanding the Dynamic Universe” funded by the Knut and Alice Wallenberg Foundation under Dnr KAW 2018.0067,  {\em Vetenskapsr\aa det}, the Swedish Research Council, project 2020-03444.

Based on observations obtained with the Samuel Oschin Telescope 48-inch and the 60-inch Telescope at the Palomar Observatory as part of the Zwicky Transient Facility project. ZTF is supported by the National Science Foundation under Grant No. AST-1440341 and a collaboration including Caltech, IPAC, the Weizmann Institute of Science, the Oskar Klein Center at Stockholm University, the University of Maryland, the University of Washington, Deutsches Elektronen-Synchrotron and Humboldt University, Los Alamos National Laboratories, the TANGO Consortium of Taiwan, the University of Wisconsin at Milwaukee, and Lawrence Berkeley National Laboratories. Operations are conducted by COO, IPAC, and UW.

This work was supported by the GROWTH project \citep{Kasliwal2019} funded by the National Science Foundation under Grant No 1545949. The SALT classification spectra of SN 2020lil and SN 2020pst were obtained through Rutgers University program 2020-1-MLT-007 (PI: S. W. Jha). Based on observations collected at the European Organisation for Astronomical Research in the Southern Hemisphere under ESO programmes 199.D-0143,
1103.D-0328, and 106.216C.009.

The authors thank Michael Tucker for advising in the quality and inclusion of SNIFS/UH-88 spectra from the SCAT survey.

Based on observations collected at Copernico 1.82m telescope (Asiago Mount Ekar, Italy) INAF - Osservatorio Astronomico di Padova. The data presented here were obtained in part with ALFOSC, which is provided by the Instituto de Astrofisica de Andalucia (IAA) under a joint agreement with the University of Copenhagen and NOT. Some of the data presented herein were obtained at Keck Observatory, which is a private 501(c)3 non-profit organization operated as a scientific partnership among the California Institute of Technology, the University of California, and the National Aeronautics and Space Administration. The Observatory was made possible by the generous financial support of the W. M. Keck Foundation. The authors wish to recognize and acknowledge the very significant cultural role and reverence that the summit of Maunakea has always had within the Native Hawaiian community. We are most fortunate to have the opportunity to conduct observations from this mountain. Based on observations obtained with the Apache Point Observatory 3.5-meter telescope, which is owned and operated by the Astrophysical Research Consortium. This article is based on observations made with the Gran Telescopio Canarias operated by the Instituto de Astrofisica de Canarias, the Isaac Newton Telescope, and the William Herschel Telescope operated by the Isaac Newton Group of Telescopes, the Italian Telescopio Nazionale Galileo operated by the Fundacion Galileo Galilei of the INAF (Istituto Nazionale di Astrofisica), and the Liverpool Telescope operated by Liverpool John Moores University with financial support from the UK Science and Technology Facilities Council. All these facilities are located at the Spanish Roque de los Muchachos Observatory of the Instituto de Astrofisica de Canarias on the island of La Palma. Based on observations collected at the European Organisation for Astronomical Research in the Southern Hemisphere. This work makes use of observations from the Las Cumbres Observatory global telescope network. Based on observations obtained at the international Gemini Observatory, a program of NSF's NOIRLab, which is managed by the Association of Universities for Research in Astronomy (AURA) under a cooperative agreement with the National Science Foundation on behalf of the Gemini Observatory partnership: the National Science Foundation (United States), National Research Council (Canada), Agencia Nacional de Investigaci\'{o}n y Desarrollo (Chile), Ministerio de Ciencia, Tecnolog\'{i}a e Innovaci\'{o}n (Argentina), Minist\'{e}rio da Ci\^{e}ncia, Tecnologia, Inova\c{c}\~{o}es e Comunica\c{c}\~{o}es (Brazil), and Korea Astronomy and Space Science Institute (Republic of Korea). Based on observations obtained at the Southern Astrophysical Research (SOAR) telescope, which is a joint project of the Minist\'{e}rio da Ci\^{e}ncia, Tecnologia e Inova\c{c}\~{o}es (MCTI/LNA) do Brasil, the US National Science Foundation's NOIRLab, the University of North Carolina at Chapel Hill (UNC), and Michigan State University (MSU). This paper includes data gathered with the 6.5 meter Magellan Telescopes located at Las Campanas Observatory, Chile. These results made use of the Lowell Discovery Telescope (LDT) at Lowell Observatory. Lowell is a private, non-profit institution dedicated to astrophysical research and public appreciation of astronomy and operates the LDT in partnership with Boston University, the University of Maryland, the University of Toledo, Northern Arizona University and Yale University. The Large Monolithic Imager was built by Lowell Observatory using funds provided by the National Science Foundation (AST-1005313). The upgrade of the DeVeny optical spectrograph has been funded by a generous grant from John and Ginger Giovale and by a grant from the Mt. Cuba Astronomical Foundation. Some of the observations reported in this paper were obtained with the Southern African Large Telescope (SALT). This work made use of the Heidelberg Supernova Model Archive (HESMA), https://hesma.h-its.org.

\end{acknowledgements}


\bibliographystyle{aa}
\bibliography{main}


\appendix
\section{Supplementary photometry}
\label{section:supplementary_photometry}

Our initial light curve quality cut was employed as a general guideline to remove any supernovae from our sample with a poor corresponding SALT fit. A total of 2954 supernovae from the \FullSampleSNe\ met this criteria. For completeness, these 2954 light curves were visually inspected, leading to the identification of 13 poorly sampled light curves which resulted in spectral phase estimates that were unrealistically early ($< -20$~d). Supplementary forced photometry for these objects was obtained from the Asteroid Terrestrial-impact Last Alert System (ATLAS; \citealt{atlas_tonry}; \citealt{atlas_smith}), leading to more reliable estimates of $t_0$.

Of the 631 objects which do not meet the initial criteria, there are a number which have acceptable enough SALT fits with respect to $t_0$ from which we can accurately estimate a spectral phase. Through visual inspection we identified 255 such objects which were then added to the sample.

For the remaining 376 targets, we scraped as many forced photometry light curves as possible from ATLAS. If an ATLAS light curve could not be located, we instead attempted to scrape photometry from the All-Sky Automated Survey for Supernovae (ASAS-SN; \citealt{asassn}). In total this search returned 197 light curves from ATLAS and 2 from ASAS-SN. We fit these supplementary light curves with SALT and inspected the fits by eye. In the case of a successful SALT fit to this additional data which produces a well constrained estimate of maximum light we `accept' the supplementary SALT fit. In total we accept the updated $t_0$ values for 84 of these objects (82 from ATLAS and 2 from ASAS-SN).

We set an errorfloor of 1.0~d on the $t_0$ value for any object that required supplementary photometry and/or did not meet the initial criteria laid out in Sect.~\ref{sec:observations}. We note here that while these objects are included in the sample at this point, the corresponding spectra may be removed subject to one of the various cuts that follow. Any spectra in the final sample corresponding to objects with these updated phases are excluded from any plots and analyses relating to light curve fit parameters other than $t_0$ ($x_1$, $c$). All updated peak MJD values coming from supplementary photometry can be found in Table~\ref{appendix:tab:updated_max_light1}.

\begin{table}
\tiny
\caption{Updated maximum light values from supplementary photometric data for improved spectral phase estimation.}
\label{appendix:tab:updated_max_light1}
\begin{tabular}{lcc}
\hline
\textbf{ZTF Name} & \textbf{Updated MJD t0} & \textbf{Photometry source} \\ \hline
ZTF18aagrebu & 58178.7$\pm$1.0 & ATLAS18mgm \\ [1.2pt]
ZTF18aaqfkqh & 58198.9$\pm$1.0 & ATLAS18mtr \\ [1.2pt]
ZTF18aaqgadq & 58249.7$\pm$1.0 & ATLAS18oin \\ [1.2pt]
ZTF18abiirfq & 58331.5$\pm$1.0 & ATLAS18sqq \\ [1.2pt]
ZTF18abywcnl & 58400.3$\pm$1.0 & ATLAS18vui \\ [1.2pt]
ZTF18acbwiic & 58426.8$\pm$1.0 & ATLAS18ycb \\ [1.2pt]
ZTF18accdxpa & 58420.6$\pm$1.0 & ATLAS18xaz \\ [1.2pt]
ZTF18accpjrx & 58412.3$\pm$1.0 & ATLAS18xlz \\ [1.2pt]
ZTF18acdfpne & 58312.6$\pm$1.0 & ATLAS18rvf \\ [1.2pt]
ZTF18acdnjqh & 58423.1$\pm$1.0 & ATLAS18znr \\ [1.2pt]
ZTF18acefgoc & 58406.3$\pm$1.0 & ATLAS18xlv \\ [1.2pt]
ZTF18achaqmd & 58450.4$\pm$1.0 & ATLAS18zek \\ [1.2pt]
ZTF18acswtoq & 58463.9$\pm$1.0 & ATLAS18zmv \\ [1.2pt]
ZTF18acwzawt & 58188.5$\pm$1.0 & ATLAS18mis \\ [1.2pt]
ZTF18acybqhe & 58335.1$\pm$1.0 & ATLAS18sss \\ [1.2pt]
ZTF18aczddnw & 58454.8$\pm$1.0 & ATLAS18zvg \\ [1.2pt]
ZTF18aczzrkn & 58488.2$\pm$1.0 & ATLAS18bcfl \\ [1.2pt]
ZTF18adazgdh & 58489.6$\pm$1.0 & ATLAS19acw \\ [1.2pt]
ZTF19aajwhse & 58526.0$\pm$1.0 & ATLAS19kkz \\ [1.2pt]
ZTF19aakpmoy & 58539.4$\pm$1.0 & ATLAS19dhw \\ [1.2pt]
ZTF19aalahqe & 58538.9$\pm$1.0 & ATLAS19dqu \\ [1.2pt]
ZTF19aanjvqr & 58222.3$\pm$1.0 & ATLAS18mzf \\ [1.2pt]
ZTF19aarnqzw & 58610.3$\pm$1.0 & ATLAS19hun \\ [1.2pt]
ZTF19aawlnlq & 58613.7$\pm$1.0 & ATLAS19jrk \\ [1.2pt]
ZTF19aawscnp & 58636.1$\pm$1.0 & ATLAS19lqt \\ [1.2pt]
ZTF19aaxpjwk & 58595.5$\pm$1.0 & ATLAS19emj \\ [1.2pt]
ZTF19aaydpru & 58635.3$\pm$1.0 & ATLAS19lgf \\ [1.2pt]
ZTF19abcttsc & 58677.2$\pm$1.0 & ATLAS19omd \\ [1.2pt]
ZTF19abdsntm & 58676.4$\pm$1.0 & ATLAS19oif \\ [1.2pt]
ZTF19abjkfrf & 58301.1$\pm$1.0 & ATLAS17odb \\ [1.2pt]
ZTF19abqgwuy & 58654.0$\pm$1.0 & ASASSN-19ol \\ [1.2pt]
ZTF19abxdtew & 58746.1$\pm$1.0 & ATLAS19uro \\ [1.2pt]
ZTF19acchtyp & 58775.1$\pm$1.0 & ATLAS19ynq \\ [1.2pt]
ZTF19aceeexa & 58779.1$\pm$1.0 & ATLAS19ykk \\ [1.2pt]
ZTF19acmyljb & 58786.4$\pm$1.0 & ATLAS19ztf \\ [1.2pt]
ZTF19acnwvuw & 58800.8$\pm$1.0 & ATLAS19zjy \\ [1.2pt]
ZTF19acrddcz & 58813.6$\pm$1.0 & ATLAS19balj \\ [1.2pt]
ZTF19acrmmpq & 58810.6$\pm$1.0 & ATLAS19bfsx \\ [1.2pt]
ZTF19actfpmy & 58814.1$\pm$1.0 & ATLAS19bapx \\ [1.2pt]
ZTF19acuymeu & 58826.0$\pm$1.0 & ATLAS19bbnc \\ [1.2pt]
ZTF19acxngol & 58823.9$\pm$1.0 & ASASSN-19acx \\ [1.2pt]
ZTF19acxolhu & 58832.9$\pm$1.0 & ATLAS19bcxj \\ [1.2pt]
ZTF19acyftrs & 58829.1$\pm$1.0 & ATLAS19bcjz \\ [1.2pt]
ZTF19acygbye & 58827.9$\pm$1.0 & ATLAS19bced \\ [1.2pt]
ZTF19acymuae & 58835.6$\pm$1.0 & ATLAS19bcub \\ [1.2pt]
ZTF19aczjsdy & 58822.8$\pm$1.0 & ATLAS19bbhj \\ [1.2pt]
ZTF19aczjyre & 58819.2$\pm$1.0 & ATLAS19bbhw \\ [1.2pt]
ZTF19adcecwu & 58863.2$\pm$1.0 & ATLAS19bfni \\ [1.2pt]
ZTF19adcghbt & 58845.1$\pm$1.0 & ATLAS19benj \\ [1.2pt]
ZTF20aacciep & 58850.8$\pm$1.0 & ATLAS19bfrh \\ [1.2pt]
ZTF20aaflngq & 58877.8$\pm$1.0 & ATLAS20bkl \\ [1.2pt]
ZTF20aagmrid & 58868.9$\pm$1.1 & ATLAS20avi \\ [1.2pt]
ZTF20aakzezp & 58862.8$\pm$1.0 & ATLAS20hfd \\ [1.2pt]
ZTF20aalkcea & 58895.4$\pm$1.0 & ATLAS20fde \\ [1.2pt]
ZTF20aamifit & 58891.4$\pm$1.0 & ATLAS20fsi \\ [1.2pt]
ZTF20aasmueu & 58916.5$\pm$1.0 & ATLAS20hwj \\ [1.2pt]
ZTF20aasoarv & 58916.3$\pm$1.0 & ATLAS20hzz \\ [1.2pt]
ZTF20aasxfcd & 58924.2$\pm$1.0 & ATLAS20ikg \\ [1.2pt]
ZTF20aatpoef & 58941.8$\pm$1.0 & ATLAS20iwi \\ [1.2pt]
ZTF20aaublmd & 58931.8$\pm$1.0 & ATLAS20ibe \\ [1.2pt]
ZTF20aaubntq & 58937.8$\pm$1.0 & ATLAS20ibk \\ [1.2pt]
ZTF20aaukudw & 58953.5$\pm$1.0 & ATLAS20jsc \\ [1.2pt]
ZTF20aavxraf & 58928.3$\pm$1.0 & ATLAS20hmy \\ [1.2pt]
ZTF20aawfnro & 58937.9$\pm$1.0 & ATLAS20ibn \\ [1.2pt]
\hline
\end{tabular}
\end{table}

\begin{table}
\tiny
\caption{Table~\ref{appendix:tab:updated_max_light1} continued.}
\label{appendix:tab:updated_max_light2}
\begin{tabular}{lcc}
\hline
\textbf{ZTF Name} & \textbf{Updated MJD t0} & \textbf{Photometry source} \\ \hline  
ZTF20aayngca & 58985.8$\pm$1.0 & ATLAS20lvx \\ [1.2pt]
ZTF20aazlvht & 58994.8$\pm$1.0 & ATLAS20mcj \\ [1.2pt]
ZTF20aazppkz & 58922.8$\pm$1.0 & ATLAS20hid \\ [1.2pt]
ZTF20abahpah & 58992.9$\pm$1.0 & ATLAS20mfn \\ [1.2pt]
ZTF20abkhhjn & 59024.8$\pm$1.0 & ATLAS20pld \\ [1.2pt]
ZTF20abkhibz & 59019.6$\pm$1.0 & ATLAS20pli \\ [1.2pt]
ZTF20abmmymj & 59061.9$\pm$1.0 & ATLAS20tia \\ [1.2pt]
ZTF20abpsdmr & 58265.5$\pm$1.0 & ATLAS18pje \\ [1.2pt]
ZTF20abqqwui & 59077.4$\pm$1.0 & ATLAS20xbd \\ [1.2pt]
ZTF20abvyauy & 59091.8$\pm$1.0 & ATLAS20xni \\ [1.2pt]
ZTF20abxyajd & 59070.0$\pm$1.0 & ATLAS20ugl \\ [1.2pt]
ZTF20acduffd & 59125.7$\pm$1.0 & ATLAS20bcef \\ [1.2pt]
ZTF20acedqji & 59131.9$\pm$1.0 & ATLAS20bcms \\ [1.2pt]
ZTF20acogywb & 58993.1$\pm$1.0 & ATLAS20lti \\ [1.2pt]
ZTF20acqhysq & 59175.5$\pm$1.0 & ATLAS20bfpg \\ [1.2pt]
ZTF20acuhjpn & 59185.0$\pm$1.0 & ATLAS20bgfa \\ [1.2pt]
ZTF20acwvmms & 59197.2$\pm$1.0 & ATLAS20bhkh \\ [1.2pt]
ZTF20acxnxbh & 59212.5$\pm$1.0 & ATLAS20bihi \\ [1.2pt]
ZTF20acyhena & 59175.4$\pm$1.0 & ATLAS20bfgx \\ [1.2pt]
ZTF20acyonmr & 59139.7$\pm$1.0 & ATLAS20bcmf \\ [1.2pt]
\hline
\end{tabular}
\end{table}

\begin{figure}[b]
\includegraphics[width = \columnwidth]{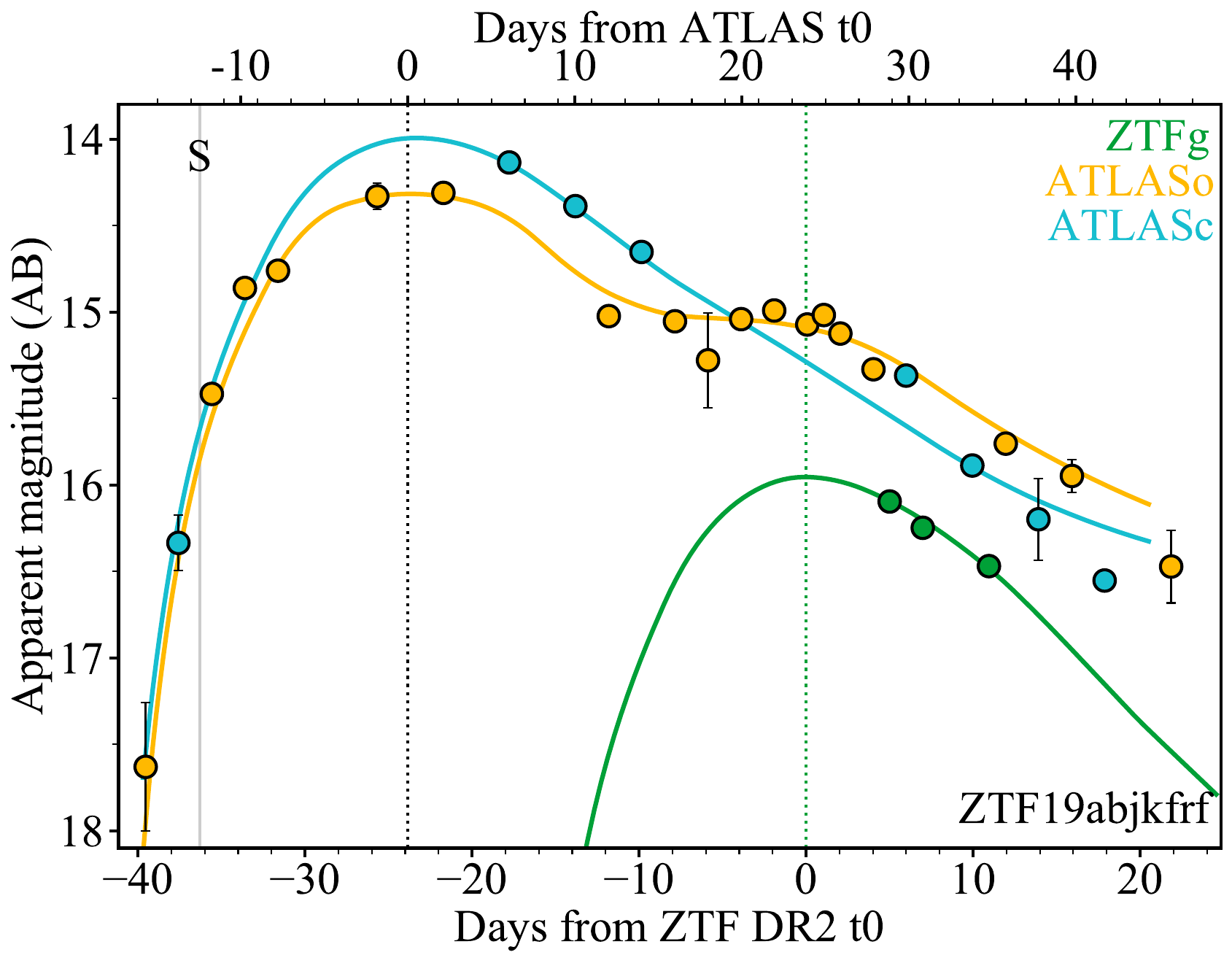}
    \caption{An example of a SALT2 fit to a DR2 \textit{g}-band light curve of ZTF19abjkfrf, which results in a poor spectral phase estimate due to low coverage and single-band data. The original SALT2 fit is shown in green with the updated fit to the supplementary data from ATLAS in the cyan (\textit{ATLASc}) and orange (\textit{ATLASo}) filters displayed by the cyan and orange curves. The spectral date is represented by the solid grey vertical line, with the dotted vertical lines corresponding to the original (green) and updated (black) maximum light fit measurements.}
    \label{appendix:fig:refit_light curve}
\end{figure}

Figure~\ref{appendix:fig:refit_light curve} illustrates one such example of an updated $t_0$ estimate. As visible in the plot, there are only three ZTF photometry points for ZTF19abjkfrf, in a single band, post-peak. Due to the poor phase coverage of the light curve, this object will be excluded from almost all studies, especially those concerning cosmological measurements. As we possess a good quality pre-peak spectrum however, updating the phase estimate in this fashion allows us to avoid cutting otherwise acceptable data. For this object the phase estimate has changed from $-$35.9 to $-$12.5~d. The updated estimates for $t_0$ for the 92 supernovae in Table~\ref{appendix:tab:updated_max_light1} were subsequently used to as initial constraints in refitting the corresponding light curves for the DR2.

\section{Data}

\textcolor{black}{We present the feature measurements for all \FinalSpectra\ spectra in Tables~\ref{appendix:tab:pv_spectra1}-\ref{appendix:tab:hv_spectra2}. For other measurements presented in this paper - lightcurve properties, Branch and Wang classifications, host galaxy properties - we refer the reader to the source papers (\citealt{georgios_dr2, burgaz_dr2_spectra}, Smith et al. in prep.).}

\begin{table*}
\tiny
\caption{Measurements of the \text{\PVFSpectra} spectra not containing a HV feature in the \SiFeature.}
\label{appendix:tab:pv_spectra1}
\centering
\begin{tabular}{lcccccccc}
\hline
\textbf{ZTF Name} & \textbf{IAU} & \textbf{MJD} & \textbf{Phase (d)} & \textbf{a$_{\text{PV}}$} & \textbf{b$_{\text{PV}}$ (\r A)} & \textbf{c$_{\text{PV}}$ (\r A)} & \textbf{pEW$_{\text{PV}}$ (\r A)} & \textbf{Telescope-Instrument} \\ \hline
ZTF19abyzuos & 2019qdj & 58753.4 & $-0.8\pm{0.2}$ & $0.240\pm^{0.005}_{0.005}$ & $6127.0\pm^{1.3}_{1.3}$ & $52.9\pm^{1.9}_{1.8}$ & $63.8\pm^{2.2}_{2.1}$ & ARC-DIS \\ [1.55pt]
ZTF18abadjhx & 2018cro & 58289.4 & $-4.0\pm{0.1}$ & $0.189\pm^{0.006}_{0.006}$ & $6109.1\pm^{2.8}_{2.8}$ & $70.7\pm^{3.9}_{3.6}$ & $66.9\pm^{4.1}_{3.7}$ & ARC-DIS \\ [1.55pt]
ZTF19acewotx & 2019sql & 58779.4 & $-2.4\pm{0.2}$ & $0.223\pm^{0.004}_{0.004}$ & $6128.7\pm^{1.2}_{1.2}$ & $54.4\pm^{1.6}_{1.5}$ & $60.7\pm^{1.8}_{1.7}$ & ARC-DIS \\ [1.55pt]
ZTF19aaywlmi & 2019hqg & 58658.3 & $-1.8\pm{0.1}$ & $0.181\pm^{0.003}_{0.003}$ & $6129.7\pm^{1.0}_{1.1}$ & $47.6\pm^{1.3}_{1.3}$ & $43.2\pm^{1.2}_{1.1}$ & ARC-DIS \\ [1.55pt]
ZTF20absvtnc & 2020rpn & 59086.3 & $-7.0\pm{0.1}$ & $0.232\pm^{0.006}_{0.006}$ & $6092.2\pm^{2.0}_{2.0}$ & $75.4\pm^{3.0}_{2.7}$ & $87.8\pm^{3.2}_{3.0}$ & ARC-DIS \\ [1.55pt]
ZTF18aazsabq & 2018crn & 58288.3 & $-5.4\pm{0.0}$ & $0.253\pm^{0.004}_{0.004}$ & $6099.6\pm^{1.4}_{1.3}$ & $74.4\pm^{1.8}_{1.8}$ & $94.3\pm^{2.2}_{2.2}$ & ARC-DIS \\ [1.55pt]
ZTF18aauizcr & 2018btb & 58257.3 & $-8.5\pm{0.1}$ & $0.293\pm^{0.005}_{0.005}$ & $6033.0\pm^{1.9}_{2.0}$ & $88.5\pm^{2.7}_{2.6}$ & $130.1\pm^{4.0}_{3.8}$ & ARC-DIS \\
\hline
\end{tabular} 
\end{table*}

\begin{table*}
\tiny
\caption{Table~\ref{appendix:tab:pv_spectra1} continued.}
\label{appendix:tab:pv_spectra2}
\centering
\begin{tabular}{lcccccccc}
\hline
\textbf{ZTF Name} & \textbf{IAU} & \textbf{MJD} & \textbf{Phase (d)} & \textbf{a$_{\text{PV}}$} & \textbf{b$_{\text{PV}}$ (\r A)} & \textbf{c$_{\text{PV}}$ (\r A)} & \textbf{pEW$_{\text{PV}}$ (\r A)} & \textbf{Telescope-Instrument} \\ \hline
ZTF18aczdetm & 2018khg & 58480.1 & $-5.4\pm{0.1}$ & $0.351\pm^{0.008}_{0.007}$ & $6113.7\pm^{1.6}_{1.5}$ & $66.6\pm^{2.2}_{2.1}$ & $117.3\pm^{3.3}_{3.1}$ & LT-SPRAT \\ [1.55pt]
ZTF20abjbiqf & 2020nxh & 59042.9 & $-2.4\pm{0.1}$ & $0.167\pm^{0.007}_{0.007}$ & $6103.7\pm^{3.6}_{3.4}$ & $65.2\pm^{4.8}_{4.5}$ & $54.5\pm^{3.7}_{3.4}$ & LT-SPRAT \\ [1.55pt]
ZTF20abeywdn & 2020mon & 59028.9 & $-2.9\pm{0.1}$ & $0.242\pm^{0.010}_{0.009}$ & $6130.1\pm^{2.3}_{2.2}$ & $56.7\pm^{3.1}_{2.9}$ & $68.9\pm^{3.3}_{3.2}$ & LT-SPRAT \\ [1.55pt]
ZTF18aaxdrjn & 2018cdt & 58272.9 & $-6.1\pm{0.1}$ & $0.295\pm^{0.009}_{0.008}$ & $6107.8\pm^{1.9}_{1.9}$ & $57.0\pm^{2.5}_{2.4}$ & $84.4\pm^{3.3}_{3.3}$ & LT-SPRAT \\ [1.55pt]
ZTF20aawmlfq & 2020igy & 58973.1 & $-4.6\pm{0.1}$ & $0.221\pm^{0.007}_{0.008}$ & $6131.0\pm^{2.6}_{2.5}$ & $57.4\pm^{3.2}_{3.1}$ & $63.7\pm^{3.4}_{3.3}$ & LT-SPRAT \\ [1.55pt]
ZTF20abciflj & 2020lhe & 59008.9 & $-3.6\pm{0.1}$ & $0.209\pm^{0.007}_{0.006}$ & $6109.2\pm^{3.3}_{3.1}$ & $83.6\pm^{4.3}_{4.3}$ & $87.6\pm^{5.0}_{4.7}$ & LT-SPRAT \\ [1.55pt]
ZTF20aalbjpf & 2020bpy & 58888.9 & $-3.2\pm{0.2}$ & $0.301\pm^{0.006}_{0.005}$ & $6111.9\pm^{1.7}_{1.7}$ & $66.4\pm^{2.5}_{2.4}$ & $100.3\pm^{3.4}_{3.3}$ & LT-SPRAT \\
\hline
\end{tabular} 
\end{table*}

\begin{table*}
\tiny
\caption{Table~\ref{appendix:tab:pv_spectra1} continued.}
\label{appendix:tab:pv_spectra3}
\centering
\begin{tabular}{lcccccccc}
\hline
\textbf{ZTF Name} & \textbf{IAU} & \textbf{MJD} & \textbf{Phase (d)} & \textbf{a$_{\text{PV}}$} & \textbf{b$_{\text{PV}}$ (\r A)} & \textbf{c$_{\text{PV}}$ (\r A)} & \textbf{pEW$_{\text{PV}}$ (\r A)} & \textbf{Telescope-Instrument} \\ \hline
ZTF20acwujgz & 2020acrz & 59205.3 & $-1.5\pm{0.1}$ & $0.336\pm^{0.007}_{0.008}$ & $6127.2\pm^{1.7}_{1.6}$ & $61.4\pm^{2.4}_{2.2}$ & $103.5\pm^{3.7}_{3.6}$ & NTT-EFOSC2 \\ [1.55pt]
ZTF18adazeua & 2018lef & 58483.3 & $-3.9\pm{0.3}$ & $0.239\pm^{0.009}_{0.009}$ & $6086.1\pm^{3.1}_{3.3}$ & $67.3\pm^{5.0}_{4.5}$ & $80.8\pm^{5.9}_{5.8}$ & NTT-EFOSC2 \\ [1.55pt]
ZTF18acurmpd & 2018awn & 58229.1 & $-7.0\pm{0.2}$ & $0.108\pm^{0.007}_{0.007}$ & $6130.8\pm^{4.3}_{4.0}$ & $60.1\pm^{6.4}_{5.7}$ & $32.4\pm^{3.5}_{3.0}$ & NTT-EFOSC2 \\ [1.55pt]
ZTF18acwzawt & 2018aby & 58186.3 & $-2.1\pm{1.0}$ & $0.324\pm^{0.005}_{0.005}$ & $6135.8\pm^{1.4}_{1.3}$ & $58.8\pm^{1.5}_{1.4}$ & $95.6\pm^{2.2}_{2.2}$ & NTT-EFOSC2 \\ [1.55pt]
ZTF18acyymoy & 2018but & 58260.3 & $-4.3\pm{0.3}$ & $0.329\pm^{0.005}_{0.005}$ & $6121.6\pm^{1.1}_{1.1}$ & $65.9\pm^{1.5}_{1.4}$ & $108.5\pm^{2.3}_{2.2}$ & NTT-EFOSC2 \\ [1.55pt]
ZTF19aaxibby & 2019hbs & 58644.4 & $-7.1\pm{0.2}$ & $0.297\pm^{0.006}_{0.006}$ & $6084.0\pm^{1.8}_{1.9}$ & $76.7\pm^{2.7}_{2.5}$ & $114.1\pm^{3.7}_{3.7}$ & NTT-EFOSC2 \\ [1.55pt]
ZTF19abywqza & 2019pil & 58730.3 & $-2.1\pm{0.2}$ & $0.159\pm^{0.005}_{0.004}$ & $6114.1\pm^{2.1}_{2.0}$ & $72.0\pm^{3.2}_{3.1}$ & $57.4\pm^{2.7}_{2.6}$ & NTT-EFOSC2 \\
\hline
\end{tabular} 
\end{table*}

\begin{table*}
\tiny
\caption{Table~\ref{appendix:tab:pv_spectra1} continued.}
\label{appendix:tab:pv_spectra4}
\centering
\begin{tabular}{lcccccccc}
\hline
\textbf{ZTF Name} & \textbf{IAU} & \textbf{MJD} & \textbf{Phase (d)} & \textbf{a$_{\text{PV}}$} & \textbf{b$_{\text{PV}}$ (\r A)} & \textbf{c$_{\text{PV}}$ (\r A)} & \textbf{pEW$_{\text{PV}}$ (\r A)} & \textbf{Telescope-Instrument} \\ \hline
ZTF19abtsrup & 2019owu & 58733.4 & $-2.5\pm{0.2}$ & $0.325\pm^{0.005}_{0.004}$ & $6119.3\pm^{1.1}_{1.1}$ & $75.2\pm^{1.6}_{1.4}$ & $122.7\pm^{2.3}_{2.1}$ & NTT-EFOSC2 \\ [1.55pt]
ZTF20aapadwx & 2020ddr & 58905.4 & $-3.5\pm{0.3}$ & $0.335\pm^{0.004}_{0.004}$ & $6120.1\pm^{0.9}_{0.9}$ & $56.1\pm^{1.0}_{1.0}$ & $94.1\pm^{1.5}_{1.5}$ & NTT-EFOSC2 \\ [1.55pt]
ZTF19abucvhz & 2019nzo & 58718.4 & $-2.6\pm{1.0}$ & $0.342\pm^{0.012}_{0.011}$ & $6123.8\pm^{2.4}_{2.4}$ & $62.7\pm^{3.6}_{3.4}$ & $107.8\pm^{5.5}_{5.5}$ & NTT-EFOSC2 \\ [1.55pt]
ZTF19ablfaey & 2019mon & 58702.2 & $-6.8\pm{0.2}$ & $0.244\pm^{0.005}_{0.005}$ & $6095.4\pm^{1.6}_{1.6}$ & $65.8\pm^{2.2}_{2.2}$ & $80.4\pm^{2.8}_{2.6}$ & NTT-EFOSC2 \\ [1.55pt]
ZTF18adbhrjs & 2018ldw & 58483.3 & $-6.5\pm{0.2}$ & $0.321\pm^{0.012}_{0.011}$ & $6109.2\pm^{2.2}_{2.2}$ & $55.3\pm^{3.4}_{3.0}$ & $89.1\pm^{4.8}_{4.6}$ & NTT-EFOSC2 \\ [1.55pt]
ZTF19aabvfwn & 2019jf & 58492.2 & $-9.3\pm{0.2}$ & $0.276\pm^{0.009}_{0.009}$ & $6104.6\pm^{2.5}_{2.5}$ & $59.9\pm^{3.0}_{2.9}$ & $82.9\pm^{4.1}_{4.1}$ & NTT-EFOSC2 \\ [1.55pt]
ZTF18acsovsw & 2018jny & 58462.3 & $-8.9\pm{0.1}$ & $0.261\pm^{0.009}_{0.009}$ & $6098.3\pm^{2.1}_{2.1}$ & $65.6\pm^{3.0}_{3.0}$ & $85.8\pm^{3.6}_{3.3}$ & NTT-EFOSC2 \\
\hline
\end{tabular} 
\end{table*}


\begingroup
\setlength{\tabcolsep}{4pt}

\begin{landscape}
\begin{table}
\tiny
\caption{Properties and measurements of the \text{\HVFSpectra} spectra containing a HV feature in the \SiFeature.}
\label{appendix:tab:hv_spectra1}
\begin{tabular}{lcccccccccccccc}
\hline\textbf{ZTF Name} & \textbf{IAU} & \textbf{MJD} & \textbf{Phase (d)} & \textbf{a$_{\text{PV}}$} & \textbf{b$_{\text{PV}}$ (\r A)} & \textbf{c$_{\text{PV}}$ (\r A)} & \textbf{a$_{\text{HV}}$} & \textbf{b$_{\text{HV}}$ (\r A)} & \textbf{c$_{\text{HV}}$ (\r A)} & \textbf{pEW$_{\text{PV}}$ (\r A)} & \textbf{pEW$_{\text{HV}}$ (\r A)} & \textbf{R} & \textbf{$\Delta v$ (km s$^{-1}$)} & \textbf{Telescope-Instrument} \\ \hline
ZTF19acgdfjt & 2019sup & 58779.4 & $-3.1\pm{0.2}$ & $0.104\pm^{0.005}_{0.005}$ & $6107.3\pm^{7.2}_{7.6}$ & $64.5\pm^{4.9}_{4.2}$ & $0.079\pm^{0.014}_{0.014}$ & $6026.6\pm^{2.8}_{2.8}$ & $33.2\pm^{3.2}_{2.3}$ & $33.8\pm^{4.0}_{3.6}$ & $13.2\pm^{3.4}_{2.9}$ & $0.40\pm^{0.19}_{0.15}$ & $4000\pm^{237}_{318}$ & ARC-DIS \\ [1.55pt]
ZTF20aafsnpp & 2020zi & 58872.4 & $-3.5\pm{0.1}$ & $0.105\pm^{0.025}_{0.026}$ & $6093.4\pm^{25.0}_{25.2}$ & $59.7\pm^{12.7}_{13.5}$ & $0.131\pm^{0.041}_{0.044}$ & $6008.3\pm^{10.3}_{8.2}$ & $41.9\pm^{6.2}_{6.9}$ & $31.9\pm^{13.4}_{13.0}$ & $27.9\pm^{12.7}_{12.3}$ & $1.00\pm^{1.41}_{0.78}$ & $4263\pm^{635}_{976}$ & ARC-DIS \\ [1.55pt]
ZTF18adaifep & 2018koy & 58485.3 & $-3.3\pm{0.3}$ & $0.304\pm^{0.017}_{0.017}$ & $6023.6\pm^{6.1}_{7.1}$ & $90.2\pm^{2.5}_{2.2}$ & $0.129\pm^{0.024}_{0.023}$ & $5927.9\pm^{4.3}_{4.4}$ & $45.7\pm^{5.1}_{6.2}$ & $137.9\pm^{9.7}_{9.3}$ & $29.8\pm^{9.1}_{9.1}$ & $0.22\pm^{0.10}_{0.10}$ & $4787\pm^{168}_{216}$ & ARC-DIS \\ [1.55pt]
ZTF18aahjafd & 2018loz & 58229.3 & $-1.2\pm{0.1}$ & $0.168\pm^{0.034}_{0.058}$ & $6134.3\pm^{29.8}_{22.8}$ & $63.1\pm^{9.6}_{11.0}$ & $0.137\pm^{0.070}_{0.059}$ & $6042.0\pm^{20.3}_{14.7}$ & $48.8\pm^{8.6}_{9.1}$ & $53.8\pm^{17.9}_{24.3}$ & $33.5\pm^{25.7}_{18.0}$ & $0.76\pm^{1.51}_{0.63}$ & $4609\pm^{382}_{607}$ & ARC-DIS \\
\hline
\end{tabular} 
\end{table} 
\end{landscape}

\begin{landscape}
\begin{table}
\tiny
\caption{Table~\ref{appendix:tab:hv_spectra1} continued.}
\label{appendix:tab:hv_spectra2}
\centering
\begin{tabular}{lcccccccccccccc}
\hline\textbf{ZTF Name} & \textbf{IAU} & \textbf{MJD} & \textbf{Phase (d)} & \textbf{a$_{\text{PV}}$} & \textbf{b$_{\text{PV}}$ (\r A)} & \textbf{c$_{\text{PV}}$ (\r A)} & \textbf{a$_{\text{HV}}$} & \textbf{b$_{\text{HV}}$ (\r A)} & \textbf{c$_{\text{HV}}$ (\r A)} & \textbf{pEW$_{\text{PV}}$ (\r A)} & \textbf{pEW$_{\text{HV}}$ (\r A)} & \textbf{R} & \textbf{$\Delta v$ (km s$^{-1}$)} & \textbf{Telescope-Instrument} \\ \hline
ZTF20abrjmgi & 2020qxp & 59075.9 & $-10.7\pm{1.0}$ & $0.221\pm^{0.008}_{0.007}$ & $6090.7\pm^{4.8}_{5.8}$ & $77.4\pm^{3.8}_{3.2}$ & $0.157\pm^{0.015}_{0.016}$ & $5990.7\pm^{2.4}_{2.3}$ & $35.1\pm^{2.7}_{2.7}$ & $85.9\pm^{6.3}_{5.4}$ & $27.7\pm^{4.5}_{4.6}$ & $0.33\pm^{0.09}_{0.10}$ & $4949\pm^{155}_{205}$ & NOT-ALFOSC \\ [1.55pt]
ZTF18aaqcozd & 2018bjc & 58249.1 & $-4.1\pm{0.1}$ & $0.311\pm^{0.008}_{0.008}$ & $6141.3\pm^{3.7}_{4.1}$ & $66.2\pm^{2.9}_{2.4}$ & $0.168\pm^{0.020}_{0.021}$ & $6052.9\pm^{2.5}_{2.4}$ & $32.9\pm^{2.5}_{2.1}$ & $103.7\pm^{5.9}_{5.6}$ & $27.7\pm^{5.1}_{4.7}$ & $0.27\pm^{0.08}_{0.08}$ & $4346\pm^{108}_{149}$ & NOT-ALFOSC \\ [1.55pt]
ZTF20acikuon & 2020wcj & 59147.3 & $-0.9\pm{0.1}$ & $0.287\pm^{0.010}_{0.010}$ & $6094.9\pm^{5.1}_{5.9}$ & $73.3\pm^{3.1}_{2.8}$ & $0.133\pm^{0.023}_{0.022}$ & $5995.4\pm^{3.8}_{3.4}$ & $42.4\pm^{3.1}_{3.1}$ & $105.7\pm^{7.1}_{7.0}$ & $28.3\pm^{6.9}_{6.5}$ & $0.28\pm^{0.11}_{0.10}$ & $4910\pm^{116}_{162}$ & NTT-EFOSC2 \\ [1.55pt]
ZTF19abkbkyb & 2018enb & 58335.2 & $-2.5\pm{0.2}$ & $0.266\pm^{0.045}_{0.112}$ & $6150.5\pm^{27.1}_{18.1}$ & $65.4\pm^{11.4}_{8.7}$ & $0.181\pm^{0.105}_{0.077}$ & $6062.7\pm^{29.0}_{12.6}$ & $41.4\pm^{17.3}_{9.4}$ & $86.8\pm^{22.8}_{35.1}$ & $37.6\pm^{42.4}_{19.4}$ & $0.55\pm^{1.18}_{0.45}$ & $4329\pm^{413}_{782}$ & NTT-EFOSC2 \\
\hline
\end{tabular} 
\end{table} 
\end{landscape}

\endgroup


\begin{figure*}
 \includegraphics[width = \linewidth]{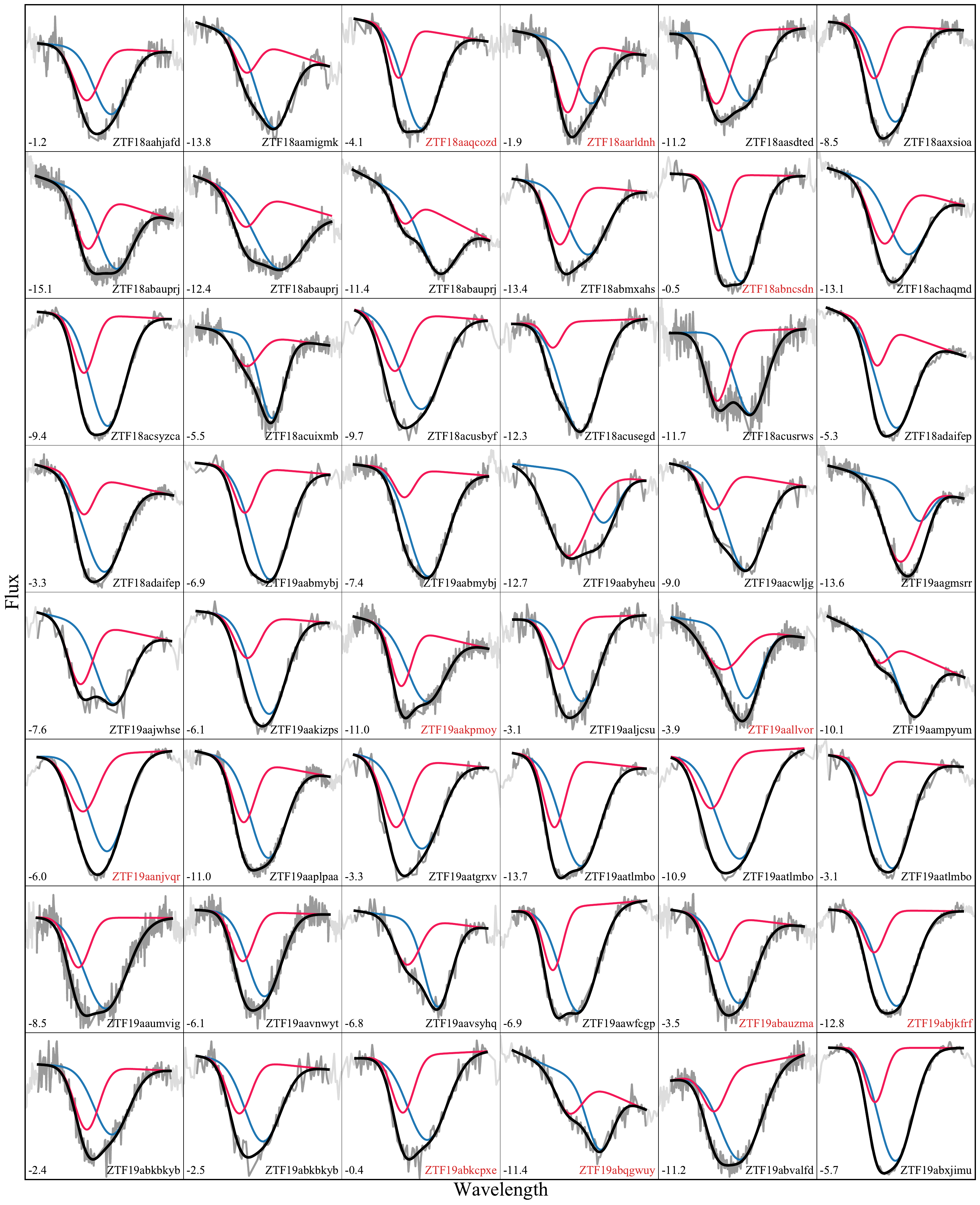}
 \caption{The best fits to the \HVFSpectra\ ZTF DR2 spectra in the sample identified as having a HV component. The phase and SN ZTF name are indicated in each panel. The continuum regions are highlighted in blue and red, the overall fits are shown by the solid black curves, and the individual PV and HV components are shown by the blue and pink curves. Red SN names indicate those that were subsequently cut in the creation of the low-bias sample.}
 \label{appendix:fig:thumnail_plots1}
\end{figure*}

\begin{figure*}
 \includegraphics[width = \linewidth]{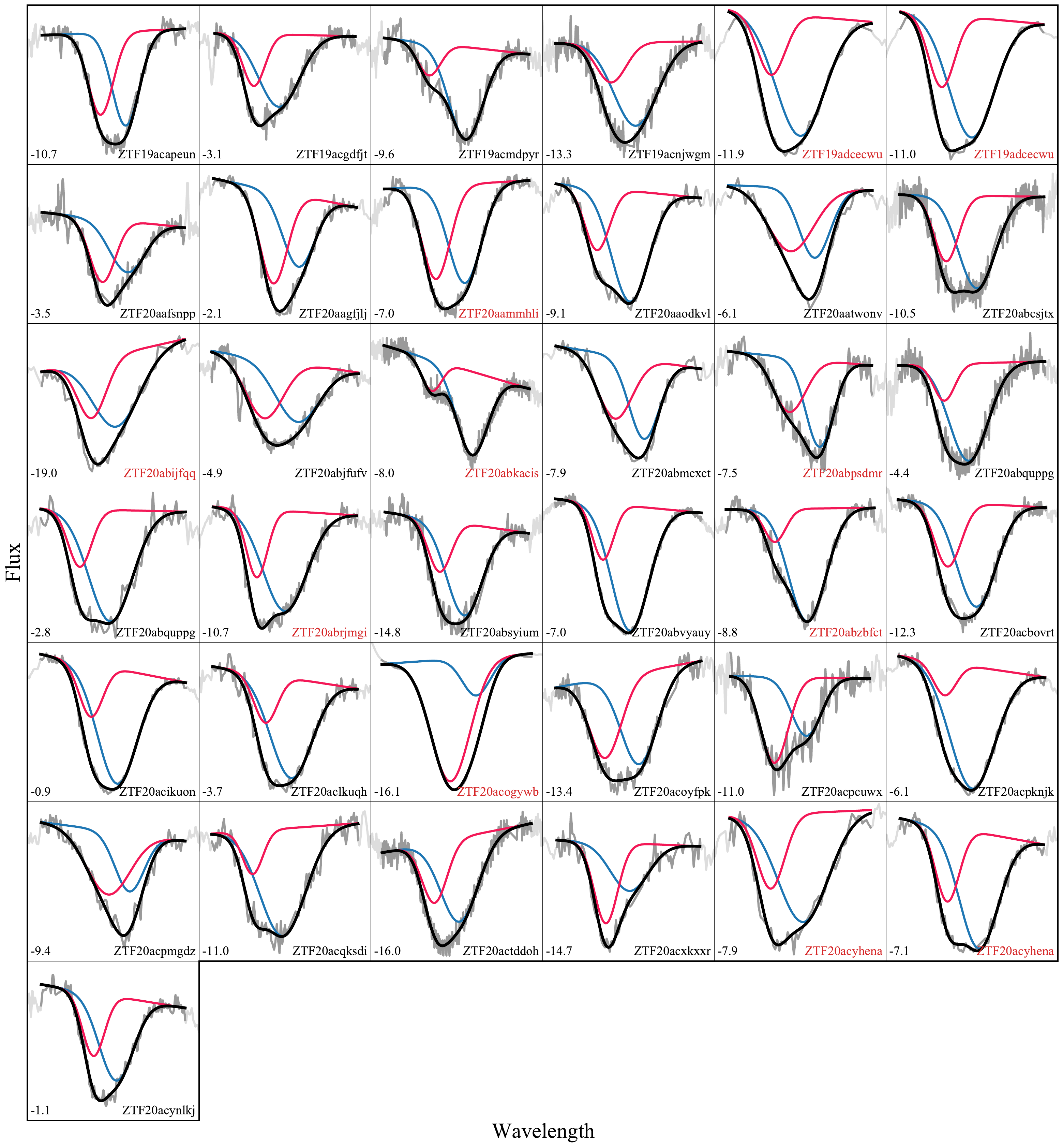}
 \caption{Continuation of Fig.~\ref{appendix:fig:thumnail_plots1}.}
 \label{appendix:fig:thumnail_plots2}
\end{figure*}

\end{CJK*}
\end{document}